\documentclass[preprint,5p,twocolumn,11pt,sort&compress]{elsarticle}

\usepackage{verbatim}
\usepackage[T1]{fontenc}
\usepackage[utf8]{inputenc}
\usepackage[american]{babel}
\usepackage{epsfig}
\usepackage{ulem}
\usepackage{graphicx}
\usepackage{booktabs}
\usepackage{multirow}
\usepackage{dcolumn}
\usepackage{amsmath}
\usepackage{mathtools}
\usepackage{amsfonts}
\usepackage{amssymb}
\usepackage{epstopdf}
\usepackage{bm}
\usepackage{siunitx}
\usepackage{braket}
\usepackage{enumitem}
\usepackage{soul}
\usepackage{xcolor}

\definecolor{darkviolet}{rgb}{0.58, 0.0, 0.83}
\usepackage{hyperref}
\definecolor{royalblue}{rgb}{0.25, 0.41, 0.88}

\newcommand{\mmt}[1]{{{ #1 }}}

\hypersetup{
    colorlinks=true, 
    linkcolor=royalblue, 
    citecolor=magenta}

\begin{document}

\begin{frontmatter}

\title{Climbing out of the shadows:\\ building the distance ladder with black hole images}

\author[Leiden]{Fabrizio Renzi}
\author[INAF]{Matteo Martinelli}

\address[Leiden]{Institute Lorentz, Leiden University, PO Box 9506, Leiden 2300 RA, The Netherlands\\ renzi@lorentz.leidenuniv.nl}
\address[INAF]{INAF - Osservatorio Astronomico di Roma, via Frascati 33, 00040 Monteporzio Catone (Roma), Italy\\ matteo.martinelli@inaf.it}

\date{\today}

\begin{abstract}
In the era of precision cosmology it has became crucial to find new and competitive probes to estimate cosmological parameters, in an effort of finding answers to the current cosmological tensions/discrepancies. In this work, we show the possibility of using observations of Super Massive Black Hole (SMBH) shadows as an anchor for the distance ladder, substituting the sources usually exploited for such purpose, such as Cepheid variable stars. Compared to the standard approaches, the use of SMBH has the advantage of not needing to be anchored with distance calibrators outside the Hubble flow since the shadows physical size can be estimated knowing the mass of the SMBH. Furthermore, SMBH are supposed to inhabit the centre of all galaxies which means, in principle, that we can measure the size of the shadows in any Supernova type Ia host galaxy. Under the assumption that the mass of the SMBH can be accurately and reliably estimated, we find that the Hubble constant can be constrained with a $\approx10\%$ precision even considering current experimental design of ground-based interferometers. By constructing a SMBH catalogue based on a specific choice of the SMBH Mass Function (BHMF), we forecast the constraints on the Hubble constant, finding that a precision of $\approx4\%$ may be within reach of future interferometers.
\end{abstract}


\end{frontmatter}


\section{Introduction}\label{sec:intro}

In present day cosmology, one of the most puzzling results is the tension in the different available measurements of the current expansion rate of the Universe, $H_0$, obtained through high and low redshift cosmological observations \cite{Abdalla:2022yfr,DiValentino:2021izs,Perivolaropoulos:2021jda,Knox:2019rjx}. Taking the most recent local measurement $H_0=73.04\pm1.04$ km s$^{-1}$ Mpc$^{-1}$ obtained by the Hubble Space Telescope (HST) and the Supernova H0 for the Equation of State (SH0ES) team \cite{Riess:2021jrx}, there is a $\approx5\sigma$ tension with the constraint inferred by the Planck Collaboration $H_0=67.4\pm0.5$ km s$^{-1}$ Mpc$^{-1}$ \cite{Planck:2018vyg}. Notice that the latter is not a direct measurement of the expansion rate, but rather a measurement of the angular size of the sound horizon at recombination, which can be related to $H_0$ by assuming that the expansion of the Universe follows the cosmological constant-cold dark matter ($\Lambda$CDM) model \cite{Planck:2018vyg}.

The former result comes instead from the "distance ladder" method, which allows to infer the value of the Hubble constant from observations of Type Ia supernovae (SNIa) immersed in the Hubble flow. The idea behind the construction of the ladder is to calibrate the absolute magnitude of cosmological SNIa, by obtaining their distance through the observation of other sources in the same system for which this can be known. The local measurement quoted above is obtained using the luminosity-period relation of Cepheids \cite{Riess:2016jrr,Riess:2019cxk,Riess:2020fzl,Riess:2021jrx} to obtain the distance of galaxies were SNIa are observed and, from this, their absolute magnitude, which can then be used with cosmological SNIa to obtain $H_0$.

Such a calibration is not the only possible, and new competitive methods have been used to anchor the distance ladder and measure the Hubble parameter, e.g. calibration using stars at the Tip of the Red Giant Branch (TRGB) from the Chicago Carnegie Hubble program (CCHP) \cite{Freedman:2019jwv,Freedman:2021ahq}, the Surface Brightness Fluctuations (SBF) method \cite{Khetan:2020hmh} or Mira variable stars \cite{Huang:2018dbn,Huang:2019yhh}. The first one in particular has reached the same accuracy as the Cepheid method, but the result shows a moderate statistical tension (2$\sigma$) with the SH0ES inference \cite{Freedman:2019jwv,Freedman:2021ahq}. Recent papers \cite{Renzi:2020fnx,Efstathiou:2020wxn,Camarena:2019rmj,Alestas:2021luu,Camarena:2021jlr,Efstathiou:2021ocp}
have shown that such a discrepancy can be related to a difference in calibrating distances to common supernovae hosts, leading to an offset of $0.2$ mag between Cepheids and TRGB SNIa calibration. Calibrations using the TRGB, independent from \cite{Freedman:2019jwv,Freedman:2021ahq}, show however very good agreement with both the SH0ES and CCHP results \cite{Yuan:2019npk,Soltis:2020gpl}, supporting the possibility that the discrepancy in calibrating SNIa distances may be due to some internal systematics of the TRGB method, e.g. in the choice of the tip of the red giant branch in stellar catalogues \cite{Yuan:2019npk}. 
In addition to these, other methods not relying on the distance ladder are also able to provide low redshift measurement of $H_0$ e.g., through   observations of Strong Lensing Time Delay \cite{Wong:2019kwg,Birrer:2020tax}, Gravitational Waves \cite{LIGOScientific:2019zcs,Leandro:2021qlc}, Quasars and Gamma-Ray Bursts \cite{Dainotti:2022bzg,Khadka:2020tlm,Khadka:2021xcc,Risaliti:2018reu}, Cosmic Chronometers \cite{Bonilla:2020wbn,Moresco:2022phi,Gomez-Valent:2018hwc} and Fast Radio Bursts \cite{Hagstotz:2021jzu}.

 While no evidence that systematics can account for the tension in the $H_0$ measurements was found \cite{Efstathiou:2019mdh,Planck:2018vyg,Riess:2016jrr,Riess:2019cxk,Riess:2021jrx,Dainotti:2021pqg,Bernal:2016gxb,DiValentino:2016hlg}, it is helpful to find new ways of calibrating SNIa measurements, in order to minimise the risk of falling under systematic effects.
 
 In this paper, we investigate the possibility of applying the distance ladder method with a completely new observable to anchor the cosmological SNIa. This relies on the possibility to use the shadow of a Super Massive Black Hole (SMBH) as a standard ruler, due to the relation existing between the size of the shadow and the mass of the SMBH \cite{Synge1966,Bisnovatyi-Kogan:2018vxl,Perlick:2018iye,Perlick:2021aok}. Such a possibility has been opened by the recent observations of Event Horizon Telescope (EHT), a telescope array consisting of a wide network of radio telescope, that for the first time has observed the immediate environment of a SMBH \cite{EventHorizonTelescope:2019dse,EventHorizonTelescope:2019uob,EventHorizonTelescope:2019ggy}. 
 
\mmt{The possible use of SMBH shadows as standard rulers was already explored in the literature \cite{Tsupko:2019pzg,Qi:2019zdk}, with the aim of constraining cosmological parameters. While our approach starts from the same idea, here we focus on the possibility of using such observations as a new calibrator for the distance ladder, thus as a viable option to obtain an estimate of $H_0$ with SNIa observations independently of the cosmological model.}

The aim of this paper is therefore to determine the constraints achievable  on the Hubble parameter by current and future interferometers dedicated to horizon-scale observations of SMBH. We also review the impact of the experimental uncertainties in the measurements of the size of the shadows and of the mass of the SMBH. We further assess how the number of observed SMBH will be affected by the angular resolution threshold of an ideal ground-based interferometer.

The paper is organised as follows. In \autoref{sec:distance_ladder} we review the main steps of the distance ladder method to obtain $H_0$ from SNIa measurements, while in \autoref{sec:BHshadow} we explore the possibility of using measurement of the angular size of SMBH shadows as a standard ruler to calibrate the SNIa data set. We produce synthetic data, as expected to be observed from the present and next generation of ground-based interferometer, following the approach of \autoref{sec:datasets} and obtain the forecasted precision on $H_0$ achievable with this method in \autoref{sec:results}. We finally summarise our conclusions in \autoref{sec:conclusions}.

\section{The distance ladder}\label{sec:distance_ladder}

The distance to any astronomical object can be written in terms of the so-called distance modulus
\begin{equation}\label{eq.distmodulus}
    \mu = m(z) - M = 5\log_{10}d_L(z) + 25\,,
\end{equation}
where $m(z)$ and $M$ are the apparent and absolute magnitude of the source while $d_L(z)$ its luminosity distance;  

The luminosity distance can be related to the line elements of the Universe given a form of the metric. Choosing a Friedmann-Lemaitre-Robertson-Walker (FLRW) line element i.e. $ds^2 = -dt^2 + a^2(t)dr^2$ (assuming $c=1$), the distance luminosity can be written in terms of the comoving distance $\chi$ as

\begin{equation}
    d_L(z) = (1+z)\int_0^z{\frac{{\rm d}z'}{H(z')}}=(1+z)\chi(z)\, ,
\end{equation}
where we have introduced the Hubble rate, $(1+z)H(z) = -dz/dt $ in terms of the redshift $z = a^{-1}(t) - 1$.

At low redshifts, \autoref{eq.distmodulus} can also be rewritten as
\begin{equation}\label{eq.absmagnitude}
    M = 5\log_{10}H_0 - 5 a_B - 25\, , 
\end{equation}
where the quantity $a_B$ is the intercept of the magnitude-Hubble relation and it is defined, for a generic expansion rate at $z>0$, with a cosmographic expansion \cite{Visser:2003vq}
\begin{multline}\label{eq.intercept}
\mathrm{e}^{a_{\mathrm B} + 0.2 m^0} = z \left[1 + \frac{1}{2}(q_0 -1) \,z \right. \\ 
    \left. - \frac{1}{6}(1-q_0-3q_0^2+j_0) \, z^2 + \mathcal{O}(z^3) \vphantom{\frac12}\right]\, ,
\end{multline}
where $m^0 = m(z\sim0)$, $q_0$ is the {\it deceleration parameter} and $j_0$ is the {\it jerk}. 

Thus, if for a given astrophysical object one has measurements of its apparent magnitude, and a way to obtain its absolute magnitude, it is possible to exploit these relations to obtain a measurement of $H_0$.

Indeed, calibrating sources to obtain such a measurement is not always feasible and therefore the sources commonly used for such an approach are "standard candles", i.e. sources whose absolute magnitude is constant.
The most common source used to infer cosmological distances are SNIa, sources that are extremely bright, and thus can be seen at quite high redshifts, and that behave as standardisable candles, i.e. there exists a relationship between SNIa peak luminosity and the shape of their light curve. While the peak luminosities are not exactly identical for all SNIa, due to the differences in mass and chemical composition of their progenitors, their intrinsic dispersion is small and therefore their absolute magnitude $M$ can be taken as constant (see e.g \cite{Pan-STARRS1:2017jku,SDSS:2014iwm,Betoule:2013abc,Mosher:2014gyd,Supercal} and reference therein for a detailed discussion about SNIa light curve reconstruction). 
Therefore, if one is able to calibrate the absolute magnitudes of SNIa at very low redshifts where their distance can be measured also by other means, such a calibration can be assumed to be valid also for SNIa that are immersed in the Hubble flow, and can therefore be used to infer $H_0$.

One method to achieve such a measurement is to identify SNIa in galaxies where Cepheids stars are present. Thanks to their period-luminosity relation, which also needs to be calibrated through observations of Cepheids in nearby hosts, one is able to obtain the distance to the host galaxy, and therefore to use this to calibrate the magnitude $M$ of the SNIa \cite{Riess:2016jrr,Riess:2019cxk,Riess:2020fzl,Riess:2021jrx}.

Connecting the distance of nearby Cepheids, with the ones in SNIa hosts and then with the SNIa in the Hubble flow is what is typically called the \textit{distance ladder} and each step of the ladder is called \textit{rung}. Practically, the distance ladder relies on a three rungs approach:
\begin{enumerate}
    \item The period-luminosity relation of Cepheids is calibrated using sources that are close enough for their distance to be measured via parallax. This allows to obtain the luminosity of more distant Cepheids;
    \item The distance of SNIa whose host galaxies also contains a Cepheid can be obtain through the period-luminosity relation of the latter. With the distance to the SNIa measured, one can use \autoref{eq.distmodulus} to obtain a measurement of $M$;
    \item with such measurement of $M$, assuming the same absolute magnitude holds for all SNIa, \autoref{eq.absmagnitude} can be used with the data of SNIa in the Hubble flow to obtain $H_0$.
\end{enumerate}

As we noted in \autoref{sec:intro}, this is not the only possible approach to calibrate the SNIa absolute magnitude. In particular, the second rung of the ladder can be performed by substituting Cepheids stars with any other methodology able to measure distance to galactic hosts: SBF, Mira variable and the TRGB are perfect examples of distance ladders independent of Cepheids distances. These alternative and independent methodologies have been proven successful in the inference of $H_0$, showing consistency within the total error budget across the various methods. However the terrific increase in accuracy that these methodologies have experienced in the latest decades have shown that some small discrepancies exist between them \cite{Freedman:2019jwv,Huang:2019yhh,Riess:2021jrx}. We further stress that the common denominator to all these methods is the use of observations that need to be calibrated with nearby distances (or {\it anchored} as it is commonly said in literature). This constitutes the first step of the ladder and the same anchors are used across the various methods.  While still within the region of being moderate statistical fluctuations, these discrepancies have led to question not only the validity of these approaches, but also the assumptions made in all of the rungs of the distance ladder \cite{Martinelli:2019krf,Efstathiou:2020wxn,Efstathiou:2021ocp,Yuan:2019npk,Camarena:2019rmj,Camarena:2021jlr,Alestas:2021luu,Renzi:2020fnx}.
 
It is therefore timely to find new methodologies that can overcome some or all these problems, in particular the need for anchoring in local hosts. 

In this work, we investigate a different method to calibrate supernovae observations, and therefore to obtain the current rate of the expansion of the Universe $H_0$, exploring the possibility of obtaining the distance to low redshift supernovae by observing the shadow of SMBHs in their host galaxies. As we elaborate later in this manuscript, SMBH shadows can be self-calibrated (if one has measurements of the mass of the BH casting the shadow) removing the need for anchoring with nearby hosts. Furthermore, SMBH (and BH generally speaking) are supposed to be present in any elliptical or spiral galaxies (and perhaps less massive and active BH with mass within $10^4$ and $10^6$ solar masses may be present in irregular and dwarf galaxies) making them very abundant in the Universe.

\section{Black Hole shadows as standard rulers}\label{sec:BHshadow}

The recent observations by EHT have highlighted how the measurement of the immediate environment of SMBHs is now within the reach of current facilities \cite{EventHorizonTelescope:2019uob,Roelofs:2019nmh,Fish:2020abc,Palumbo:2019abc,Haworth:2019urs}. Such observations can provide significant information since, when surrounded by an emission region, black holes are expected to exhibit a dark shadow caused by effects due to gravitational lensing of light rays and by the capture of photons at the event horizon \cite{Falcke:1999pj,Bronzwaer:2021lzo}. The shadow is caused by the difference in luminosity between the accelerated particles falling into the BH and the BH itself (which is a perfect absorbing surface or else a perfect black body at temperature $T\sim 0 $\footnote{It is worth noting that the BH surface is expected to emit a thermal radiation by Hawking mechanism \cite{Hawking:1975vcx,Visser:1997yu,Visser:2001kq}; however such a radiation would have an energy so small that it would be negligible with respect to the luminosity of the photons orbiting around the BH.}). The apparent angular size of the shadow can be obtained from the size of the cone of photon escaping \cite{Synge1966}
\begin{equation}\label{eq:angsize}
    \sin^2\theta_{\rm BH} = \left(1 - \frac{2m}{r}\right)\left( \frac{3\sqrt{3}m}{r}\right)^2\,,
\end{equation}
where we assumed a Schwarzschild BH, defined the reduced mass as $m = G M_{\rm BH}/c^2$ and $ r $ is the distance between the observer and the centre of the BH. For a distant observer one can take the limit $r \gg 2 m $ and promote the radial coordinate of the Schwarzschild metric into the angular diameter distance $d_A$ of the FRLW metric (see e.g. \cite{Perlick:2018iye,Bisnovatyi-Kogan:2018vxl} for a full derivation). This leads to (assuming $\sin \theta \approx \theta$)
\begin{equation}\label{eq.BH_apparentsize}
    \theta_{\rm BH}(z) \approx \frac{l_{\rm BH}}{d_A^{\rm BH}(z)}\,,
\end{equation}
where $l_{\rm BH}=3\sqrt{3}m$ is the physical size of the shadow.
BH shadows can be therefore used as standard rulers if one is able to obtain $l_{\rm BH}$, i.e. a measure of the BH mass, $M_{\rm BH}$. This allows to infer the distance of a BH by measuring, independently, the apparent angular size of the shadow for a system at redshift $z$, $\theta_{\rm BH}(z)$, and the mass $M_{\rm BH}$ of the SMBH. 

\mmt{It is important to stress here that until this point we assumed that the BHs under examination can be described by the Schwarzschild metric. However, such an assumption cannot hold for all BHs, since when the spin of the BH is non-vanishing, one needs to use metrics able to describe rotating BHs, such as the Kerr metric \cite{Kerr:1963ud}. 
Throughout the rest of this paper, we keep relying on the Schwarzschild description; we assess the impact of considering Kerr BHs on our approach in \ref{app.KerrBHs}.}

\mmt{In addition to the correct description of the spacetime around the BHs, several other complications might affect the modelling that leads to \autoref{eq:angsize} \cite{Vagnozzi:2020quf}; evolving accretion mechanisms and changes in the plasma distribution of the accretion disk can indeed lead to modifications of this equation \cite{Perlick:2015vta,Chowdhuri:2020ipb}, and second order effects on the photons travelling to the observer, such as weak gravitational lensing, can also have an impact. In this work, we neglect these complications, taking advantage of the fact that we are only using BH shadows as calibrators for the distance ladder. This allows us to focus the analysis on systems at low redshift, where the details of the BH are better understood and modelled, while the effect of weak gravitational lensing is negligible, due to its nature of an integrated effect along the line of sight.}

The recent observations by EHT highlighted how the relation between $\theta_{\rm BH}$, $l_{\rm BH}$ and $M_{\rm BH}$ can exploited to measure with great precision the mass of the BH, having measurements of its distance and of the apparent angular size of the shadow \cite{EventHorizonTelescope:2019ggy,EventHorizonTelescope:2022xnr}.

In this work, we take a complementary approach, and investigate instead the possible use of such observations for cosmological inference provided one is able to measure the mass of the SMBH with other methods.
We show that, under these assumptions, the observation of shadows of SMBHs is able to provide a measurement of the distance between these systems and the observer.

Knowing the black hole mass and the angular size of the shadow on the sky, it is indeed possible to infer the angular distance of the BH in a similar fashion to what is done for Baryon Acoustic Oscillations (BAO), which, similarly, rely on the observation of a standard ruler, i.e. the size of the sound horizon at recombination, to infer angular distances \cite{Ross:2014qpa,DESI:2016fyo}. Compared to BAO, however, the BH shadows have the advantage of not being related to recombination physics, but only rely on  the ability to measure the mass of the BH independently from the measurement of the size of the shadow , and on the validity of the assumed theory of gravity (General Relativity in this work). There is yet another advantage, the BAO peak in the galaxy 2D correlation function has amplitude that decreases as $z\rightarrow 0$ so that in our local Universe the peak is suppressed by non-linear fluctuations produced by peculiar velocities or other bulk effects making difficult to measure the BAO angular scale below $z\lesssim 0.05$ \cite{Ross:2014qpa,DESI:2016fyo} . Conversely, the BH shadows angular size grows linearly with the BH mass which, in turn, is expected to be maximal today since the BHs have had time to grow accreting mass from the stars and gas that surround them or by merging with other BHs \cite{Kormendy:2013dxa,King:2016abc}. This peculiarity of the BH evolution allows us to measure bigger shadow sizes in our local Universe than at cosmological distance and open the possibility of using the BH shadows to perform distance measurements at low redshifts. 

In the following section we describe in detail how measurements of $d_A(z)$ could be achieved by upcoming observations; however, in order to use such measurements as a calibrator for the distance ladder, we still need to convert the angular distance of the shadow ($d_A^{\rm BH}$) into a luminosity distance ($d_L^{\rm BH})$. Throughout the paper, we will assume that the mutual scaling of the two distances is fixed by the reciprocity theorem $d_A(1+z)^2 = d_L$; such a relation implicitly assumes that the Distance Duality Relation (DDR) holds, an assumption that could be violated in non-standard cosmological models \cite{EUCLID:2020syl,Hogg:2020ktc,Renzi:2020bvl,Renzi:2021xii}. However, departures from this relation are a cumulative effect, and in the low redshift regime, where we perform our investigation, the DDR holds at least at first approximation. This allows to write the luminosity distance to the BH shadows as: 
\begin{equation}\label{eq.BH_lumdist}
    d_L^{\rm BH} = (1+z)^2\frac{3\sqrt{3}G}{c^2}\frac{M_{\rm BH}}{\theta_{\rm BH}}\,.
\end{equation}

\section{Supernovae and Black Holes data sets}\label{sec:datasets}

In order to forecast the results of the distance ladder calibrated with upcoming measurements of SMBH shadows, we need to simulate the data sets that will be available in the near future. We need therefore to create data for both SNIa and observations of SMBH shadows. We take
\begin{itemize}
    \item mock catalogues for future observations of SMBH shadows, where we impose a cut in the angular size of the shadow corresponding to the angular resolution of EHT and post-EHT experiments;
    \item mock SNIa data for the Legacy Survey of Space and Time (LSST) of the Vera C. Rubin observatory \cite{LSST:2008ijt}.
\end{itemize}

We provide details on how these data sets are built in \autoref{sec:BHdata} and \autoref{sec:sndata} respectively.

\subsection{A catalogue for SMBH shadows}\label{sec:BHdata}
In this section we describe how to create our synthetic realisation of a SMBH shadows catalogue compatible with upcoming astrophysical observations. 
We start from the local SMBH distribution to get a catalogue of events with an associated mass and redshift. Given the redshift and assuming a cosmological model, we can obtain the angular diameter distance for each of these events, while from \autoref{eq:angsize} and \autoref{eq.BH_apparentsize} we can obtain both the intrinsic and apparent size of the shadow for each of the events. We then cut out of the catalogue the events for which the shadow cannot be resolved with the chosen experimental set up, thus introducing a minimal observable size $\theta_{\rm cut}$, and we associate an error on the observables ($M_{\rm BH}$ and $\theta_{\rm BH}$) for each of the events.

\subsubsection{The SMBH mass function}

The first ingredient we need for our catalogue is the distribution in redshift and mass of the SMBHs that can be observed. This can be obtained starting from the local BH mass function (BHMF), which is the number of SMBH per unit volume and mass
\begin{equation}
    \phi_{\rm BH} = \frac{dN}{dVd\log M}\, .
\end{equation}

Obtaining this distribution from observations is not trivial, as astronomical surveys are inevitably incomplete. This is mostly due to selection bias in the counts of SMBH in any mass bin and because there is yet no consensus on a broadly applicable and precise technique to determine accurately the mass of the SMBH \cite{Kelly:2012abs}. Thus many different variants of the BHMF exists nowadays.  \cite{Salucci:1998yp,Marconi:2003tg,Aller:2002rp,Greene:2007xw,Natarajan:2008ks,Lauer:2006jg,Kelly:2013abc}

As commonly done in literature \cite{Graham:2007qb,Tucci:2016tyc,Merloni:2008hx}, we choose here a Schecter-like BHMF with functional shape
\begin{equation}\label{eq:phibm}
     \phi_{\rm BH}(M) = \phi_\star\left(\frac{M}{M_\star}\right)^{1+\alpha}\exp\left({1 - \frac{M}{M_\star}}\right)\, ,
\end{equation}
where $\phi_\star$ is a normalisation constant and $M_\star$ is the cut-off mass of the BH population.

Tuning the parameters of the BHMF we can represent very different BH populations and assess the impact of different parameterizations on the final population of SMBH. We choose here to use a representation of the BHMF as derived in \cite{Shankar:2007zg,Shankar:2013dwa}. The parameters of such BHMF are derived phenomenologically fitting observations, obtained inferring the SMBH masses from the bolometric luminosity of early and late-type galaxies with the functional form above. The fit is performed in the log-range of BH masses $\log_{10}M_{\rm BH} \in [6,9]$ in the local Universe ($z \sim 0$).   This BHMF can be obtained fixing the parameters to $\alpha = -1.19$ , $\log_{10} M_\star = 8.4$, and $\log_{10}\phi_\star = -3 $ \cite{Tucci:2016tyc}. Consequently, this BHMF is characterised by an exponential cut-off for masses above  $M_\star \sim 10^9 M_\odot $. However it has been argued that ultra-massive BH with masses above $10^9$ do exists in nature, but a cutoff may still exist preventing SMBH to have arbitrary high masses \cite{Natarajan:2008ks,King:2016abc}. To account for this, we use a different value for $M_\star$ that we fix corresponding to the estimates of \cite{Pesce:2021adg} to $M_\star = 3.5\times 10^{10} \ M_\odot$.
Such a value is obtained calibrating the estimates in \cite{Natarajan:2008ks} for the upper limit of the mass of local SMBHs with the observational constraints on the mass of the SMBH in NGC600 \cite{Thomas:2016aaa}. This pushes the exponential cutoff of both mass functions to masses around $10^{11}\ M_\odot$.
We will refer to this model as SH09MX.

Until now, we have discussed the local BHMF; however, the SMBHs we observe today have evolved from primordial seeds by (primarily) accreting matter from their surroundings and by merging. While the knowledge of accretion mechanisms have evolved significantly in the past decades \cite{Merloni:2008hx,Valiante:2017abc,King:2008au,Inayoshi:2019fun}, a well motivativated physical model that can describe the evolution of SMBH throughout cosmic history is far from being conceived and one has typically to rely on phenomenological models to describe the mass evolution of SMBH (see e.g \cite{Cavaliere:1971abc,Small:1992abc}). 

It is therefore clear that the BHMF needs to evolve in time, thus in order to asses the number of SMBH of a given mass at different cosmological epochs one needs to model such an evolution. We choose here to follow a fully phenomenological approach that is based on few simple assumptions to construct a "visibility function", reproducing the evolution of the BHMF in a simplistic way. While we do not expect it to be as accurate as a model which takes into account the complicated mechanisms involved in SMBH accretion, we will show that it gives realistic prediction given the uncertainties in the modelling of the SMBH mass function.

The assumptions that we use to construct our visibility function are the following:

\begin{itemize}
	
	\item the conditional probability to observe a SMBH with mass $M_{BH}$ at redshift $z$, $P(M_{\rm BH}|z)$, is described by a log-Gaussian distribution, and, therefore, the distribution in $\log M$ will be Gaussian;
	\item the higher mass tail of the BHMF has evolved from seeds with mass of the order $ 10^5 - 10^6 \ M_\odot$, thus we assume this is the lower limit of the mass for the BH we consider;
	
	\item  we assume that the scaling in redshift of the mean and variance of the BHMF follows the scaling of distances in FLRW, $ M_{\rm BH} \propto R_{\rm BH} \propto (1+z)^{-1} $.
	 
\end{itemize}

Implementing these three conditions we can describe the redshift evolution of the BHMF by combining the local mass function with the conditional probability at a given redshift.

The first assumption allows us to write the conditional probability of measuring a mass $M_{\rm BH}$ at redshift zero as a Gaussian distribution in $x = \log_{10}M_{BH}$. 
The variance of this distribution is fixed requiring that at 2 standard deviation the maximum mass measurable is $ 10^{11}\ M_\odot $, in agreement with what we have discussed previously about the existence of a cutoff mass for ultra-massive BHs. This consideration can be translated in an equation to determine the standard deviation of the Gaussian $\sigma_M$ by imposing \cite{Renzi:2020bvl}
\begin{equation}
	\int^{11}_{5} P(x|z=0)dx - 0.95 = 0 \,.
\end{equation}
This assumption would still allow for the presence of ultra-massive BH with $M_{\rm BH}>10^{11}\ M_\odot$ with a probability of $5\%$. In other words, the high mass tail of the distribution would populate our catalogue with ultra-massive BH with masses that can exceed significantly the threshold of $10^{11}\ M_\odot$. 
However, it must be taken into account that when combining with the BHMF, this would weight the higher mass tail of the distribution through an exponential cutoff which prevents SMBHs to have masses much bigger than $M_\star$.

Applying the second condition means that at high redshift a considerable number of SMBH still exist with mass of the order of $ 10^5 - 10^6\ M_\odot $. This translate into an higher probability of measuring these masses compared to masses higher than $ 10^6\ M_\odot $. Combining this with the third condition, implies that only the variance of the distribution of $P(x)$ will scale with redshift while its mean remains constant.  

The final distribution for the mass $M_{\rm BH}$ in logspace will be constructed normalising the peak of the distribution to unity at all redshift. The form of this distribution is as follows:
\begin{equation}\label{eq.conditional}
    P(x|z)= \mathcal{N}\left(x=5,\frac{\sigma_M}{1+z}\right)\, .
\end{equation}
This avoids that at $z\sim 0$ the distribution deviates significantly from the local BHMF and at higher redshift the distribution would only shrink towards small masses rather than changing also in amplitude. In other words as we increase the value of $z$ the distribution will scale self-consistently. Notice that, in order to exactly recover the local BHMF at $z=0$, one would need to assume a uniform distribution i.e. $P(x|z=0) = 1$, which would return the local BHMF when the two are combined. However, we find that our assumption of a Gaussian distribution leads to negligible deviations as it impacts only the higher mass tail of the BHMF.

Multiplying this distribution with the local BHMF of \autoref{eq:phibm}, $\phi_{\rm BH}$, we obtain our final expression for the BHMF at given mass and redshift
\begin{equation}\label{eq.BHMF}
	\Phi(x,z)  = \phi(x)P(x|z)\, .
\end{equation}

It is worth stressing that for the purposes of this work, using SMBH shadows to calibrate low redshift SNIa, we only need a catalogue of local SMBH, which are described by $\phi(x)$. However, the redshift distribution $\Phi(x,z)$ is necessary to compute the total number of systems, which will directly impact the number of systems that fall in the redshift range that we will use to calibrate SNIa. Indeed, a different choice of $P(x|z)$ could impact the results and affect the final outcome of our methodology. While we rely on this simple approach throughout the rest of this work, we also assess the impact of such an assumption in \ref{app:redscaling}.

\subsubsection{Simulated observations}

With the redshift dependent BHMF now defined, we can generate our synthetic SMBHs catalogue, i.e. a set of mock observations of mass and redshfit. To do so, we apply the same procedure used to realize synthetic catalogues of gravitational waves observations (see e.g. \cite{Zhao:2010sz,Belgacem:2018lbp}) and integrate $\Phi$ across the chosen range of mass and redshift. 
By construction, the integral of $\Phi$ across a range of mass and redshift gives the number of available SMBH in the comoving volume:
\begin{equation}\label{eq.norm}
    N_{\rm tot} = 4\pi A \int_5^{11}  dx \int_0^6 dz \frac{\chi^2(z)}{H(z)}\Phi(x,z)\, ,
\end{equation}
where $A$ is a normalisation constant and we have assumed a standard FLRW background to express the comoving volume in terms of the comoving distance $\chi(z)$ and the Hubble parameter $H(z)$. 

From the above equation, it is clear that we can easily normalise the BHMF to unity and use it as a probability density to extract the mass and redshift needed for our SMBH catalogue once the total number of SMBH observables in the Universe is fixed. 
\mmt{The estimates of \cite{Pesce:2021adg} show that the number of SMBHs exceeding the observational threshold of shadows with size $\theta> 10\ \mu$as and flux $F> 10^{-3}$ Jy and having optically thin disks (i.e. for experiment like EHT) is $\mathcal{O}(10)$. We found that fixing $N_{\rm tot} = 10^6$ our catalogue has only $6$ shadows that exceed the threshold of $ 10\ \mu$as in agreement with the estimates of \cite{Pesce:2021adg}. We therefore fix the total number of observables BHs (conservatively) to $N_{\rm tot}=10^6$. 
This will be our baseline setting for the total number of SMBHs. We stress however that the estimate of  \cite{Pesce:2021adg} falls short in predicting the number of M87-like shadows of a factor of $\sim 50$ possibly making our estimates over pessimistic; we will assess in \autoref{sec:results} how changing $N_{\rm tot}$ affects the final results.}

Having fixed the normalisation through \autoref{eq.norm} we can finally write the bidimensional distribution for mass and redshift as
\begin{equation}
P(x,z) = \frac{4\pi A}{N_{\rm tot}} \frac{\chi^2(z)}{H(z)}\Phi(x,z)\, ,
\end{equation} and the masses and redshifts of the SMBHs in our catalogue can then be obtained performing a MCMC sampling of $P(x,z)$. 

Following this approach, we end up with a set of N pairs of $\{M_{\rm BH},z_{\rm BH}\}$; we then assume a fiducial $\Lambda$CDM cosmology, with $H_0=70$ km s$^{-1}$ Mpc$^{-1}$ and $\Omega_{\rm m}=0.29$, which allows to associate a distance $d_A^{\rm BH}$ to each redshift. Using \autoref{eq.BH_apparentsize} we can then obtain the apparent angular size of the shadow for each mock SMBH.
The catalogue is therefore transformed in a set of observations $\{M_{\rm BH},\theta_{\rm BH}\}$ for each event drawn. Of the full catalogue we only keep the events at low redshift ($z\lesssim 0.01$), i.e. the range in which we will calibrate the SNIa. Finally, to each event left we associate observational errors that we assume as $\sigma_\theta/\theta_{\rm BH}=0.07$ \cite{EventHorizonTelescope:2019dse}, and $\sigma_M/M_{\rm BH}=0.07$ \cite{Gebhardt:2011abc,Oldham:2016abc,Walsh:2013uua}. \mmt{This assumption on the relative uncertainties of the observables can be extremely optimistic, in particular for what concerns the measurement of the BH mass. While for the rest of our analysis we assume the values above, we also assess the impact of less optimistic assumptions on the final constraints in \ref{app:uncertainties}.}

Notice that the choice of the starting local BHMF can significantly impact  the distribution of masses in the catalogues and determines how many of the simulated shadows will be actually observable.

\begin{figure}
    \centering
    \includegraphics[width=0.9\columnwidth,keepaspectratio]{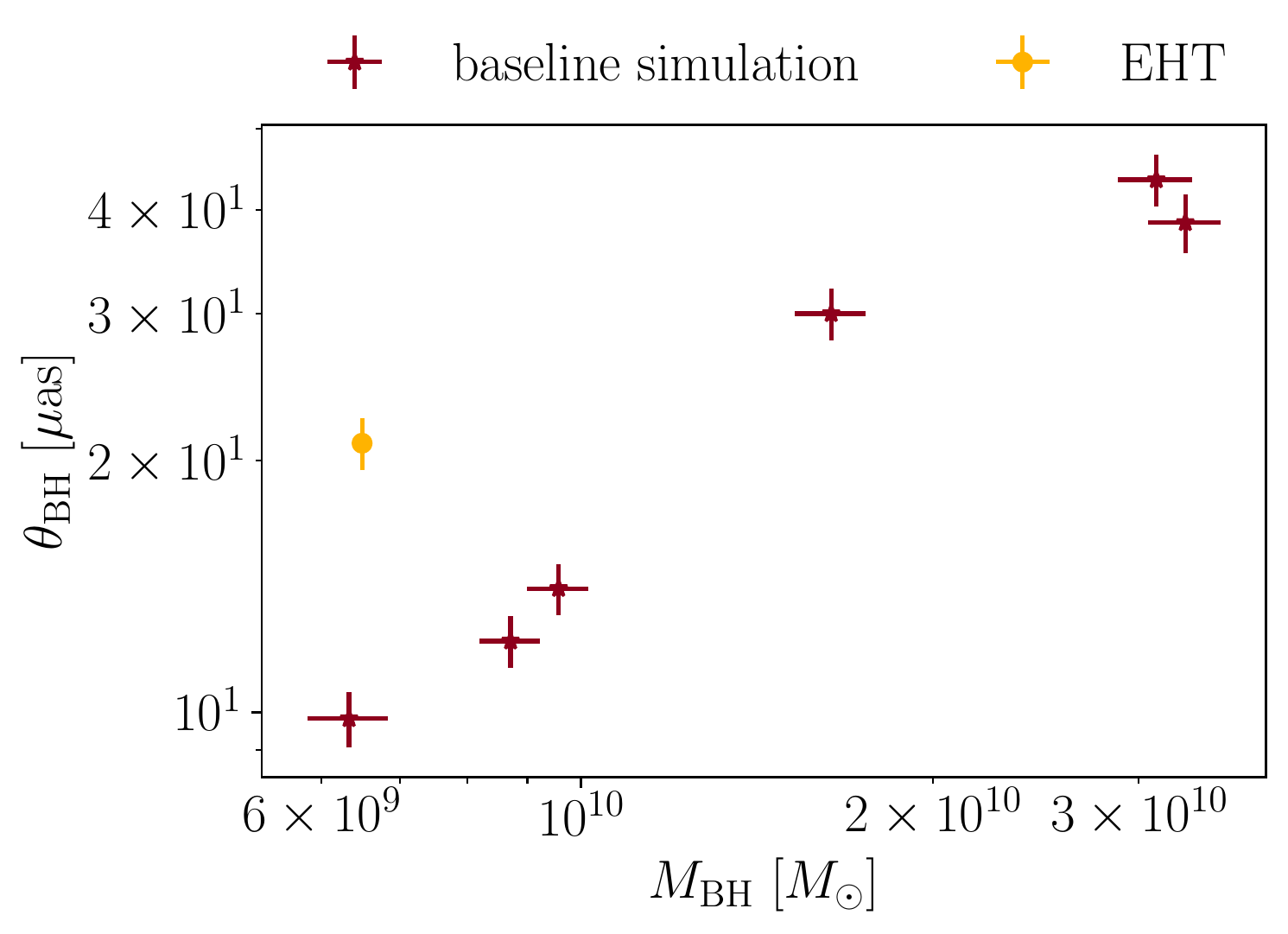}
    \caption{The SMBH shadows baseline catalogue obtained from the SH09MX mass function. We report here with yellow data points only the SMBH appearing below $z \leq 0.01$ and for $\theta_{\rm cut} = 10 \mu$as. The red data point shows the event observed by EHT.} 
    \label{fig.catalogues}
\end{figure}

We show the final baseline catalogue in \autoref{fig.catalogues}, and we recap the assumptions made to produce it in \autoref{tab:baseline}.

\begin{table}[h]
	\centering
	\renewcommand{\arraystretch}{1.5}
	\begin{tabular}{c@{\hspace{1 cm}}c}
		 \toprule
		 \multicolumn{2}{c}{Cosmology}  \\
		 \hline
		 $\Omega_{\rm m}$  & $0.29$\\
		 $H_0$ [km/s/Mpc] & $70.0$\\
		 \hline
		 \hline
		 \multicolumn{2}{c}{Mass function}  \\
		 \hline
		 local BHMF  & SH09MX\\
		 $N_{\rm tot}$  & $10^6$\\
		 $M_{\rm cut}$ [$M_\odot$] & $10^{11}$\\
		 \hline
		 \hline
		 \multicolumn{2}{c}{Survey specifications} \\
		 \hline
		 $\theta_{\rm cut}$ [$\mu$as] & $10$\\
		 $\sigma_\theta/\theta_{\rm PBH}$ & $0.07$\\
		 $\sigma_M/M_{\rm PBH}$ & $0.07$ \\
		 \bottomrule
	\end{tabular}
	
	\caption{Choices of cosmological parameters, BHMF specifications and measurement errors used to construct our baseline catalogue.}\label{tab:baseline}
\end{table}

It is worth noting that, the large range of masses for the SMBH population makes the angular size of the shadow span several orders of magnitude even at similar redshifts. However the state-of-the-art of BH shadows observations \cite{EventHorizonTelescope:2019dse,Pesce:2021adg} only allows to observe angular sizes above a certain threshold size $\theta_{\rm cut}$. For the EHT instrument $\theta_{\rm cut} = 10\ \mu $as  \cite{EventHorizonTelescope:2019dse,Pesce:2021adg} leaving most of the SMBH available in the comoving volume invisible. In order to account for this sensitivity threshold, we apply a hard cut to our simulated catalogue, only preserving events for which $\theta_{\rm BH}>\theta_{\rm cut}$. We will however assess in \autoref{sec:results} how changing this cut can affect the results.

The events shown in \autoref{fig.catalogues} show those that survive both the angular and redshift cut (in red), alongside the data point from available EHT observations (yellow).

\subsection{SNIa catalogue}\label{sec:sndata}

As discussed above, a catalogue of SMBH shadow measurements will not be enough to constrain $H_0$ using the distance ladder. We also need SNIa observations that can be calibrated with SMBH measurements, together with a set of SNIa in the Hubble flow. The making of this additional data set requires again to estimate the redshift distribution in order to produce a mock catalogue. We start here from estimating the number distribution of SNIa from the Pantheon catalogue (which contains 1048 of such observations) \cite{Pan-STARRS1:2017jku}. This is done by reconstructing the redshift distribution of SNIa from the sources of the Pantheon catalogue and then applying an inverse transform sampling technique\footnote{Inverse transform sampling is a technique that allows to translate a uniform distribution into a generic one using its cumulative distribution function (CDF). This can be done solving a integral equation of the form : 
$$ {\rm CDF}(x) - u = \int_{-\infty}^{+ \infty}P(x) dx - u  = 0 $$ where $u$ is a sample of the uniform distribution $U[0,1]$. Solving this equation for different values of $u$ allows to extract samples of a generic probability distribution P$(x)$  } to obtain a statistically equivalent version of the Pantheon data set in redshift space. We employ the sampled redshifts to create the set of mock observations needed for the inference of $H_0$ through the distance ladder. The values of the relative magnitudes at each redshift is then calculated using \autoref{eq.distmodulus} and assuming the same fiducial cosmology employed for the SMBH catalogues. 

For this work, we fix the number of resampled SNIa to be $ 1000$ in the range $z \in [0.01,10]$ and the value of the intercept $a_B = 0.715$ corresponding to its $\Lambda$CDM value \cite{Riess:2016jrr}.
This range of redshift is used to ensure that the distribution of the SNIa in the data set is consistently reconstructed by the inverse sampling algorithm. However, the high-redshift tail of the distribution does not significantly exceed $z \sim 2$ and basically no SNIa are found above $z \sim 3 $ as expected from current observations. As a further check, we note that out of 1000 resampled SNIa we found $\sim 270$ below $z \leq 0.15$ consistent with the real Pantheon data set \cite{Pan-STARRS1:2017jku}. This ensures that our synthetic SNIa data set has exactly the same statistical properties of the observed one.

To each SNIa relative magnitude we associate an error due to the brightness uncertainties following \cite{Astier:2014swa}
\begin{equation}
    \sigma[\mu(z_s)]^2=\delta \mu(z_s)^2+\sigma^2_{\textrm{flux}}+\sigma^2_{\textrm{scat}}+\sigma^2_{\textrm{intr}}\,,
\end{equation}
where $\sigma_{\rm flux} = 0.01$ is the systematic uncertainties related to flux calibration , the intrinsic scatter of SNe at fixed colour is $\sigma_{\rm scat} = 0.025$, $\sigma_{\rm intr} = 0.12$ is the intrinsic distance scatter  and 
we include an intrinsic dispersion in the distance modulus of the form: $\delta\mu(z_s) = e_M z_s$ with $e_M$ drawn from a Gaussian distribution $\mathcal{N}(0,0.01)$ \cite{Astier:2014swa}. We use this error for all the SNIa in the second and third rungs of the ladder.

\section{Constraints on the Hubble constant}\label{sec:results}

In this section we provide details on our analysis pipeline, which exploits the method of \autoref{sec:distance_ladder} using the observables of \autoref{sec:datasets}. We then provide the constraint one can obtain on $H_0$ with both current and forecast data.

\subsection{Analysis method}\label{sec.methodology}
We start noting that the distance ladder performed with SMBH shadows is only composed of two rungs, precisely the second and third (see \autoref{sec:distance_ladder}), as SMBH shadows can provide calibrated distance measurements if the mass of the BH is known.
To perform the second rung, a pair observations of a black hole shadow and of a SNIa in the same galactic host is required. If such configurations exist, one can estimate $M$ from each pair as
\begin{align}
    M = m^{\rm SNe} - 5\log_{10} d_L^{\rm BH} = 5\log_{10}H_0 - 5a_B\, ,
\end{align}
where the above relation implicitly assumes $d_L^{\rm BH} = d_L^{\rm SNe}$. For each SMBH shadow in our catalogue, we assume that a measurement of a SNIa is available in the same galaxy, and we translate the SMBH distance, $d_L^{\rm BH}$, into an estimate of $M$. We then combine all the inferred $M$ into a joint likelihood that we employ to constrain cosmological parameters.

The second step of our ladder consists in fixing $a_B$. This can be obtained fitting the low redshift part of the cosmological SNIa catalogue that we discussed in  \autoref{sec:sndata} with the cosmographic expansion of \autoref{eq.intercept}. 

The full pipeline combines these two rungs in a simultaneous fit of the five free parameters $\{M,H_0,a_B,q_0,j_0\}$, where the absolute magnitude $M$ is used to fit the magnitude measurements inferred from the SMBH catalogue, while the remaining four enter the expressions for the luminosity distance, \autoref{eq.distmodulus} and \autoref{eq.absmagnitude}, which are used to fit the low redshift SNIa. We employ a Gaussian $\chi^2$ likelihood for $a_B$ and we combine this with the joint likelihood for $M$ to obtain our final constraints on the five parameters. These parameters are sampled using the public sampler \texttt{Cobaya} \cite{Torrado:2020dgo}, which employs a Metropolis-Hastings algorithms to sample the parameter space \cite{Lewis:2013hha,Lewis:2002ah}, assuming flat priors on the parameters set. The final results are obtained analysing the samples using \texttt{GetDist} \cite{Lewis:2019xzd}.

\subsection{Constraints using current and forecast data}

We now apply the analysis method we presented to the baseline catalogues for SMBH and SNIa generated following \autoref{sec:BHdata} and \autoref{sec:sndata}. This will provide us with forecast results for upcoming observations of multiple SMBH images and estimate the precision that one will be able to achieve on measurements of the Hubble constant. With the baseline settings of \autoref{tab:baseline}, we find that our catalogue contains six observables shadows for $z\lesssim0.01$, i.e. observations available to anchor the distance ladder.

In addition to this, we also assess the precision that could be achieved with the currently available observations, i.e. a single shadow observed with the sensitivity of EHT. In order to do so, we use the EHT measurement of $\theta_{\rm BH}=21\pm1.5$ $\mu$as \cite{EventHorizonTelescope:2019dse}\footnote{Let us note that in \cite{EventHorizonTelescope:2019dse} it is reported a measure of the size of the observed diameter of the shadow of the SMBH in M87. Therefore in the main text we reported this result and the corresponding uncertainty divided by two. However one can instead multiply by two both side of \autoref{eq:angsize} to obtain the angular size of the shadow diameter. Note also that there is a small asymmetry in the photon ring due to the SMBH rotation as discussed in \cite{EventHorizonTelescope:2019pgp}}. The EHT collaboration used such a measurement alongside an independent estimate of the distance, to obtain a measurement of the SMBH mass. Here instead we assume that an independent estimate of the mass is available, with a measurement precision of $7\%$. Furthermore, we simulate the presence of an observed SNIa in the same host of the SMBH, whose magnitude is obtained from our fiducial cosmology and with an uncertainty following \autoref{sec:sndata}. We name such a calibration data set, containing a single shadow and a single SNIa, EHT-like.

\begin{figure*}
    \centering
    \includegraphics[width=.7\textwidth,keepaspectratio]{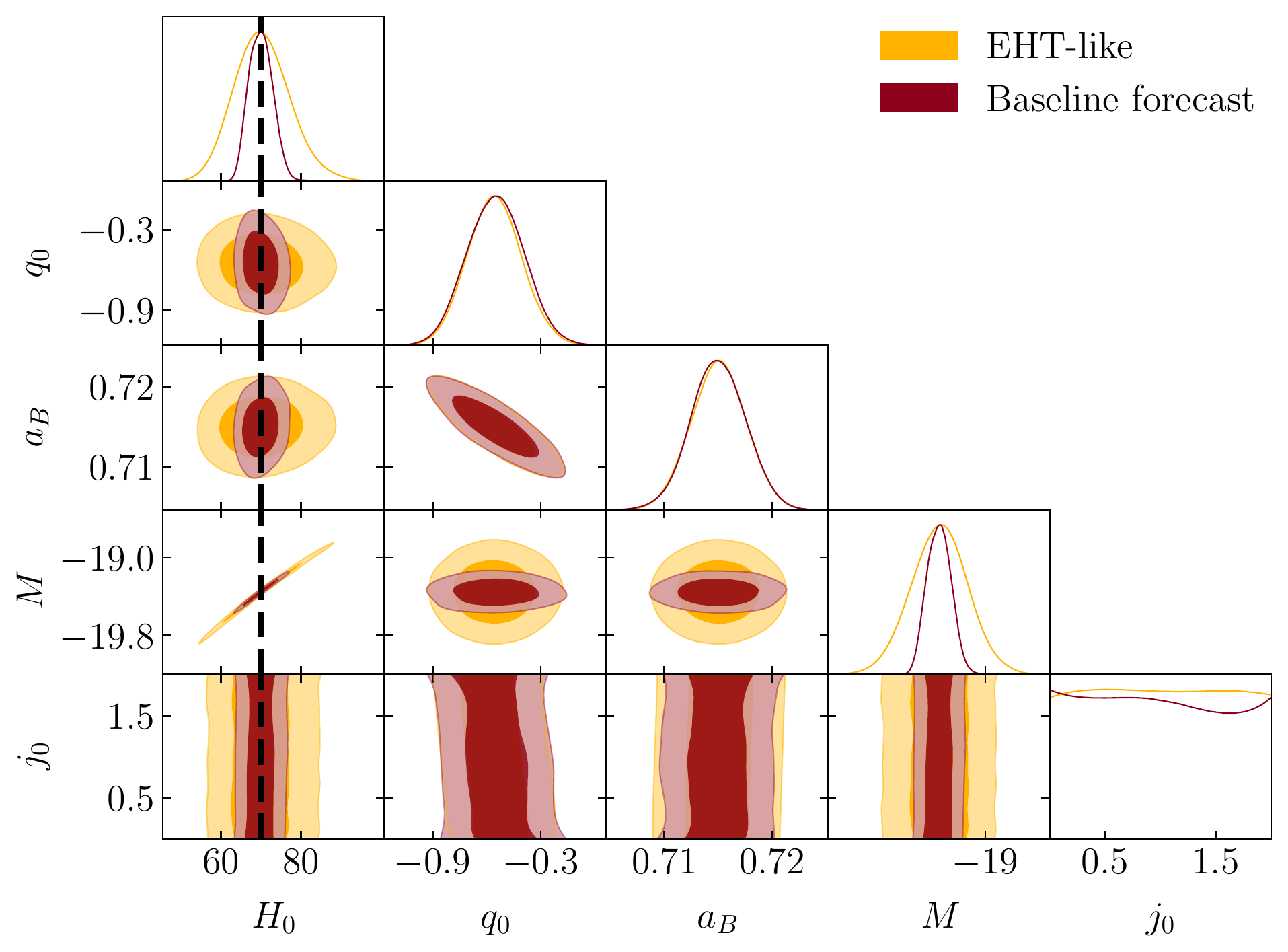}
    \caption{$68\%$ and $95\%$ confidence intervals for the free parameters of the analysis using the EHT-like (yellow) and baseline (red) data sets. The black vertical line shows the fiducial value of $H_0$ used to produce the BH and SNIa catalogues.}
    \label{fig:baseline}
\end{figure*}

We show the results obtained with EHT-like and baseline data sets in \autoref{fig:baseline}, and we report the constraints on the free parameters in \autoref{tab:results}. From these results we can see that with observations equivalent to those currently available, one can already obtain a measurement on the Hubble constant, with $H_0$ constrained with a precision of $\approx10\%$. In the baseline forecast case, with only a handful more images available, the precision on $H_0$ improves to $4\%$, i.e. a bound larger than those obtained with common methods to anchor the ladder, but still tight enough to be used to obtain independent measurements of $H_0$.

\begin{table}
	\centering
    \renewcommand{\arraystretch}{1.5}
    \begin{tabular}{c|cc}
		\toprule
		       & EHT-like             & Baseline forecast\\
		\hline
		 $H_0$ & $70.3^{+6.4}_{-7.5}$ & $69.9^{+2.7}_{-3.3}$\\
		 $a_B$ & $0.7150\pm 0.0026$   & $0.7150\pm 0.0026$\\
		 $q_0$ & $-0.56\pm 0.15$      & $-0.55\pm 0.16$\\
		 $M$   & $-19.35\pm 0.22$     & $-19.353\pm 0.091$\\
		 $j_0$ & ---                  & --- \\
 		\bottomrule  
	\end{tabular}
	
	\caption{Mean values and $68\%$ errors on the free parameter of our analysis, obtained using the EHT-like and baseline data sets (\autoref{tab:baseline}).}\label{tab:results}
\end{table}

Our results show how the bounds on $a_B$ and $q_0$ do not change switching from EHT-like to the baseline data set. This is due to the fact that the constraining power on $a_B$ and $q_0$
is mostly determined by the low redshift end of the SNIa in the Hubble flow, and they are therefore independent from the uncertainty of the calibrator distances. The uncertainty on $H_0$ is instead determined by the measurement error achievable on the absolute magnitude $M$, which is directly related to the errors on the observables, $\theta_{\rm BH}$ and $M_{\rm BH}$. In \autoref{tab:results} and \autoref{fig:baseline}, we can also notice how the jerk parameter $j_0$ is not constrained by the analysis. This is due to the fact that $j_0$ enters the cosmographic expansion at the second order in redshift; this means that its impact will be relevant at higher redshift with respect to $q_0$. Since the redshift range of the cosmological SNIa we use for our analysis is limited to $z<0.15$ (see \autoref{sec:sndata}), we do not reach an high enough redshift for $j_0$ to be relevant and therefore we are not able to constrain it.

\mmt{As mentioned in \autoref{sec:BHdata}, in our baseline settings, we assumed that the observational error on both the angular size of the BH ($\sigma_{\theta_{\rm BH}}$) and on the its mass ($\sigma_{M_{\rm BH}}$) are $7\%$ of the measured value for these observables. These assumptions on $\sigma_{M_{\rm BH}}$ and $\sigma_{\theta_{\rm BH}}$ are quite optimistic given that the observation of SMBH shadows is still far from becoming a mature field for cosmological inference \cite{Vagnozzi:2020quf}. We therefore assess the impact of less optimistic assumptions on the observational errors in \ref{app:uncertainties}, finding that this can significantly impact the final results of our approach.}



Finally, we want to stress that a constant percentage error over a whole data set is highly unrealistic; each measurement of a SMBH shadow is independent from the others and the peculiarities of each system can make shadows very different across the catalogue. The examples proposed in this manuscript would therefore corresponds to a situation in which systematic uncertainties of the SMBH are subdominant compared to the experimental errors.

\subsection{Angular threshold and total number of BHs}

\begin{figure*}
    \centering
    \begin{tabular}{cc}
    \multicolumn{2}{c}{
    \includegraphics[width=.8\textwidth,keepaspectratio]{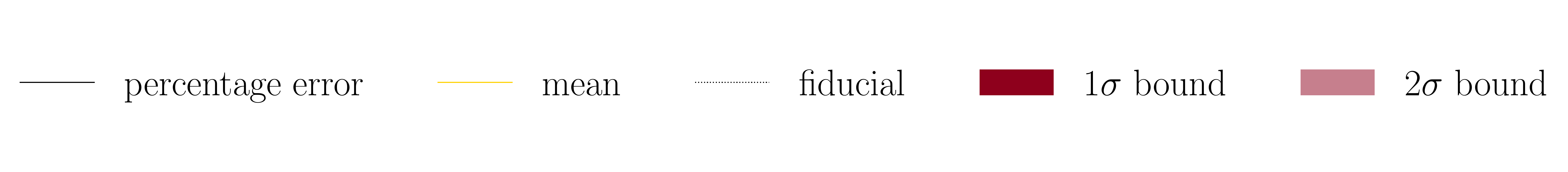}
    }\\
    \includegraphics[width=.9\columnwidth,keepaspectratio]{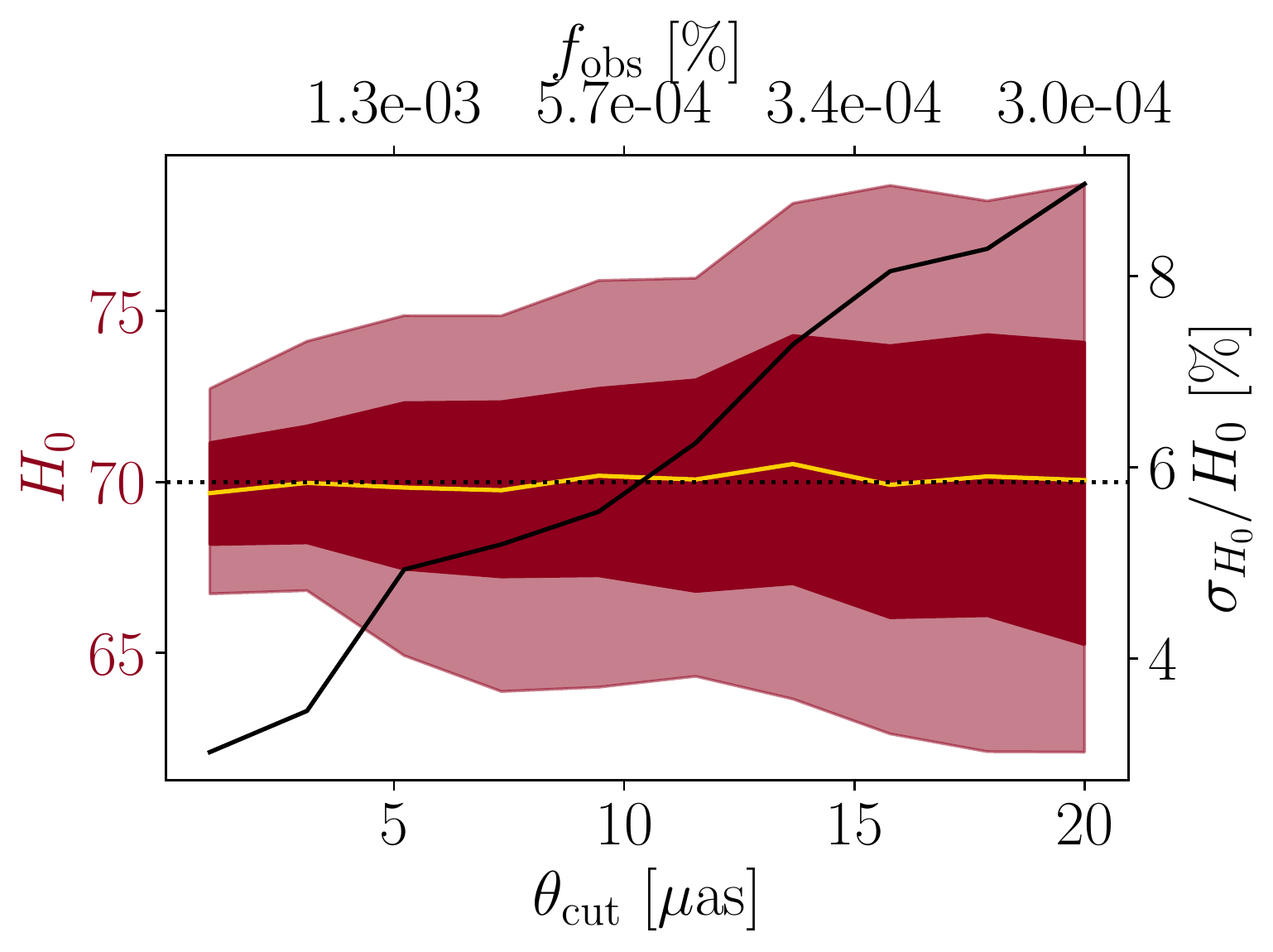} &  
    \includegraphics[width=.9\columnwidth,keepaspectratio]{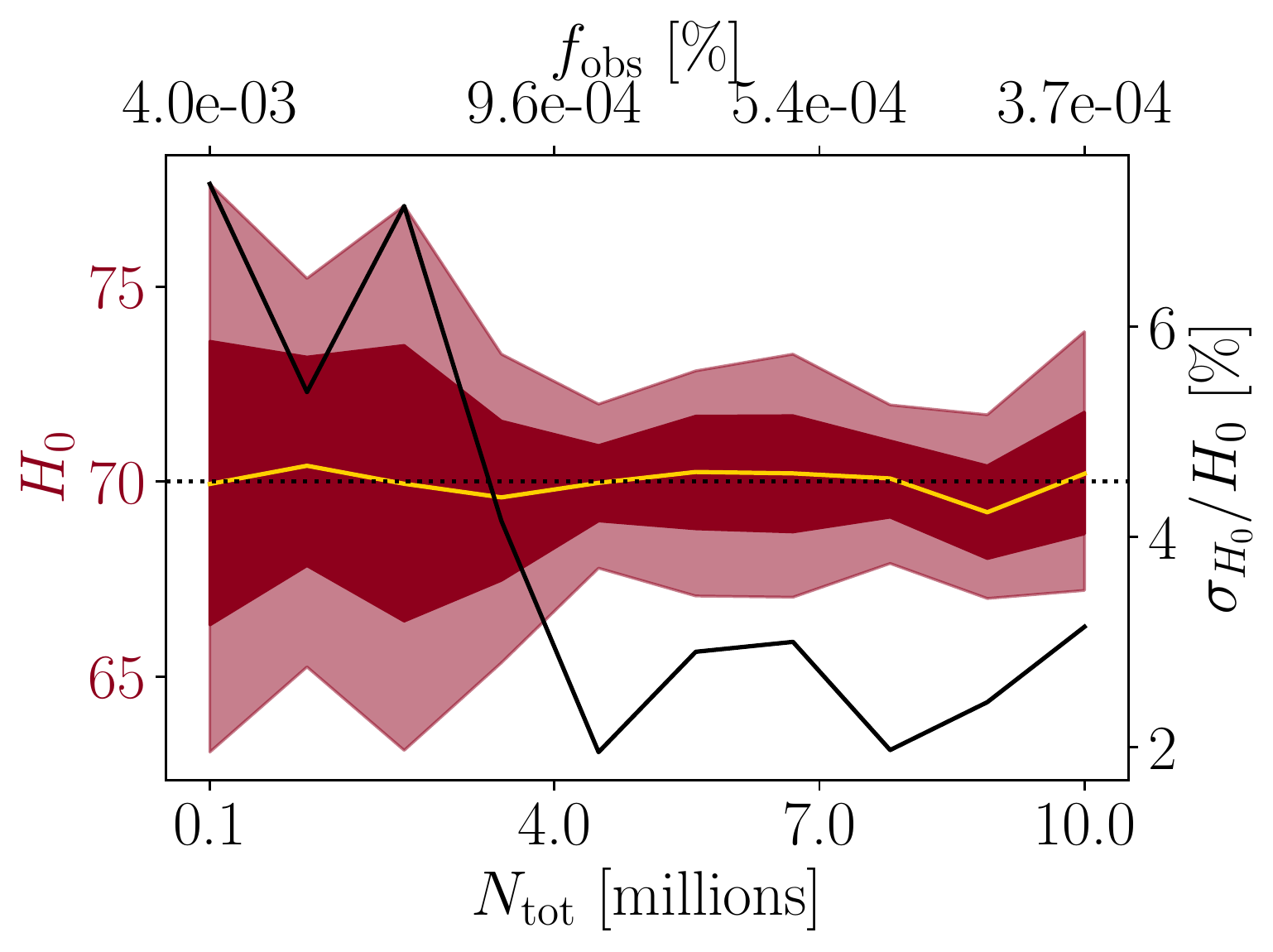}\\
    \end{tabular}
    \caption{Mean values (yellow) and $68\%-95\%$ bounds (red) for $H_0$ obtained changing the threshold observable angle $\theta_{\rm cut}$ (left panel) and the total number of SMBH $N_{\rm tot}$ (right panel). The upper x axis shows the fraction of observed events with respect to the total number of available SMBH, for each of the explored cases.}
    \label{fig:assumption_change}
\end{figure*}

While we saw above that changing the errors on the observables with respect to our baseline choices can significantly degrade our constraints, we have two other assumptions that can instead impact the results: the angular size threshold for detection ($\theta_{\rm cut}$) and the total number of expected BHs ($N_{\rm tot}$). Both these parameters will change (with respect to our baseline) the number of SMBH shadows that can be employed for the distance ladder, but in a slightly different way. The value of $N_{\rm tot}$ will determine the available SMBH per bins of mass and redshift, i.e. the number of observable shadows. The value of $\theta_{\rm cut}$ will instead determine how many of those would be actually measured by an experiment with given angular resolution. We show the bounds obtained on $H_0$ changing these parameters in \autoref{fig:assumption_change}, with the left panel referring to changes in $\theta_{\rm cut}$ and the right panel showing the changes due to $N_{\rm tot}$. As expected, an increase in $\theta_{\rm cut}$ leads to a lower fraction of the existing SMBH being observables ($f_{\rm obs}$), which translate into looser bounds on $H_0$. For the range of $\theta_{\rm cut}$ explored here, a significant change in the precision of this method is found, with our extreme cases being $\sigma_{H_0}/H_0(\theta_{\rm cut}=1\ \mu{\rm as})\approx 2\%$ and $\sigma_{H_0}/H_0(\theta_{\rm cut}=20\ \mu{\rm as})\approx 9\%$. Similar results are found when changing $N_{\rm tot}$, which directly affect the number of BH available at the redshifts of interest. We find in this case that the precision on $H_0$ can vary between $\sigma_{H_0}/H_0(N_{\rm tot}=10^{5})\approx 7\%$ and $\sigma_{H_0}/H_0(N_{\rm tot}=10^7)\approx 2\%$.

\section{Conclusion}\label{sec:conclusions}

One of the most pressing issue in modern cosmology is that of understanding the nature of cosmological tensions. A promising avenue is represented by the possibility that these discrepancies may be related to the failure of the standard cosmological model, thus unveiling the opportunity of discovering new physics beyond it. In this manuscript we have proposed a new way of performing the distance ladder with cosmological SNIa using horizon scale observations of SMBH as distance calibrator. Compared with the usage of Cepheid variable stars, SMBH shadows have the advantage of having absolute calibration (the physical size of the shadow) fixed by the theory of General Relativity (see also \autoref{sec:BHshadow}).

This feature of SMBH shadows allows us to perform the distance ladder with only two rungs; in other words the BH shadows do not require being anchored with nearby calibrator like in the case of Cepheid stars. Furthermore SMBH are thought to inhabit any galaxy in the Universe thus, provided that the shadow and the SMBH mass can be measured independently, any SNIa measurement would have an associated SMBH calibrator. 

As a proof of concept for the proposed pipeline, we applied it to mock data sets of SMBH and SNIa. First, see \autoref{sec:datasets}, we generated a synthetic realisation of a SMBH shadow catalogue bounded on a specific choice of the BHMF that we produced using a Schecter-like functional form, also employing a toy-model for the conditional probability of finding a SMBH with a given mass at a given redshift. We then imposed a cut in redshift ($z_{\rm BH}$) to keep only the SMBH present in the redshift range used to calibrate the distance ladder, and a cut $\theta_{\rm BH}>\theta_{\rm cut}=10\ \mu$as to remove from the catalogues all those SMBH shadows whose angular size is below the observation threshold. Second, we generated a mock data set for SNIa, ensuring that this is statistically equivalent to the observed Pantheon catalogue.

With these two data sets in hand, we applied the distance ladder methodology we described in \autoref{sec:distance_ladder} and \autoref{sec:BHshadow} to forecast the precision it could achieve on a measurement of $H_0$. We found that our approach could provide a precision of $\approx4\%$ on $H_0$ with the catalogues obtained with our baseline settings. We also quantified the impact of other assumptions we made during the production of the synthetic data set, finding how more pessimistic assumptions on the errors on $\theta_{\rm BH}$ and $M_{\rm BH}$, or more stringent requirement on the threshold size $\theta_{\rm cut}$, could worsen the expected precision.

\mmt{Overall,} our results seem to indicate that the approach presented in this work could be used as an alternative method to measure $H_0$, and that it can therefore contribute to the investigation of the Hubble rate tension by providing an independent measure of this parameter. Nevertheless, the precision on $H_0$ that we estimated may be in the reach of upcoming observations if a reliable method to infer the mass of the SMBH is found. Currently, as discussed in \autoref{sec:BHshadow}, there is no agreement in how to determine reliably the SMBH mass, which is critical to determine the distance of SMBHs directly from their resolved shadows. 

\mmt{The results presented in this work rely on the assumption that our choice of the metric describing the gravitational field around the SMBHs holds. Such an assumption is necessary to obtain the relation between the size of the shadow, the mass of the BH and its distance from the observer. Our baseline results are obtained assuming a Schwarzschild metric, and we investigated the effects on the results of a different assumption by switching to a Kerr metric (\ref{app.KerrBHs}); however, several other possibilities are available (see e.g. \cite{Perlick:2021aok, Amarilla:2015pgp}) and, potentially, observations of BH shadows can be used to obtain constraints on the possible metrics describing the SMBH environment, an investigation that the EHT collaboration has started, finding consistency with the Kerr metric \cite{EventHorizonTelescope:2020qrl,EventHorizonTelescope:2021dqv,EventHorizonTelescope:2022xqj}.}

 \mmt{In concluding, while the observation of SMBH shadows is still in its infancy, and many systematic effects need to be better understood in order to obtain robust and reliable constraints, the future of these measurements looks very promising and may lead to a golden era for the study of the very foundations of gravity and of our Universe.}

\section*{CRediT authorship contribution statement}
\textbf{Fabrizio Renzi:} Conceptualization, Methodology, Software, Formal analysis, Validation, Writing -- original draft, Writing - Review \& Editing. \textbf{Matteo Martinelli:} Conceptualization, Methodology, Formal analysis, Validation, Writing -- original draft, Writing - Review \& Editing.

\section*{Declaration of competing interest}

The authors declare that they have no known competing financial interests or personal relationships that could have appeared to influence the work reported in this paper.

\section*{Acknowledgements}
FR acknowledges support from the NWO and the Dutch Ministry of Education, Culture and Science (OCW), and from the D-ITP consortium, a program of the NWO that is funded by the OCW. MM acknowledges funding by the Agenzia Spaziale Italiana (ASI) under agreement n. 2018-23-HH.0.

\appendix

\section{Impact of redshift scaling on the BHMF}\label{app:redscaling}

In constructing our BHMF and its scaling with redshift we have assumed to know the form of the conditional probability $P(x|z)$ appearing in \autoref{eq.BHMF}. However, as we noted already in the main text, this assumption is quite simplistic as it does not take into account all the complicated processes involved in SMBH accretion physics (see e.g \cite{Tucci:2016tyc,Pesce:2021adg}). We therefore review in this appendix the impact of this assumption over the behaviour of our catalogue  for our baseline model. 

Let us start discussing the effect of our assumption on the evolution of the BHMF by first looking at the behaviour of the marginalised distributions of mass and redshift of the BHMF, which affect the features of the SMBH in our catalogues. These functions can be obtained simply integrating over either the mass $M_{\rm BH}$ or the redshift $z$, i.e.
\begin{align}
    P(z) &= \frac{4\pi A}{N_{\rm tot}} \frac{\chi^2(z)}{H(z)} \int^{12}_6 dx\  \Phi(x,z) \\
    P(x) &= \frac{4\pi A}{N_{\rm tot}}  \int_0^6 dz\ \frac{\chi^2(z)}{H(z)}\Phi(x,z)
\end{align}
For the conditional probability we choose instead a model with an additional parameter describing the index of the scaling with redshift i.e.

\begin{equation}\label{eq.condP_scaling}
    P(x|z) = \mathcal{N}(x=5,\sigma_M (1+z)^p)
\end{equation}

and we choose two values of $p$ corresponding to our baseline model i.e. $p = -1 $ and a value corresponding to $p_1 = -0.8$. 

\begin{figure*}
    \centering
    \includegraphics[width=\columnwidth,keepaspectratio]{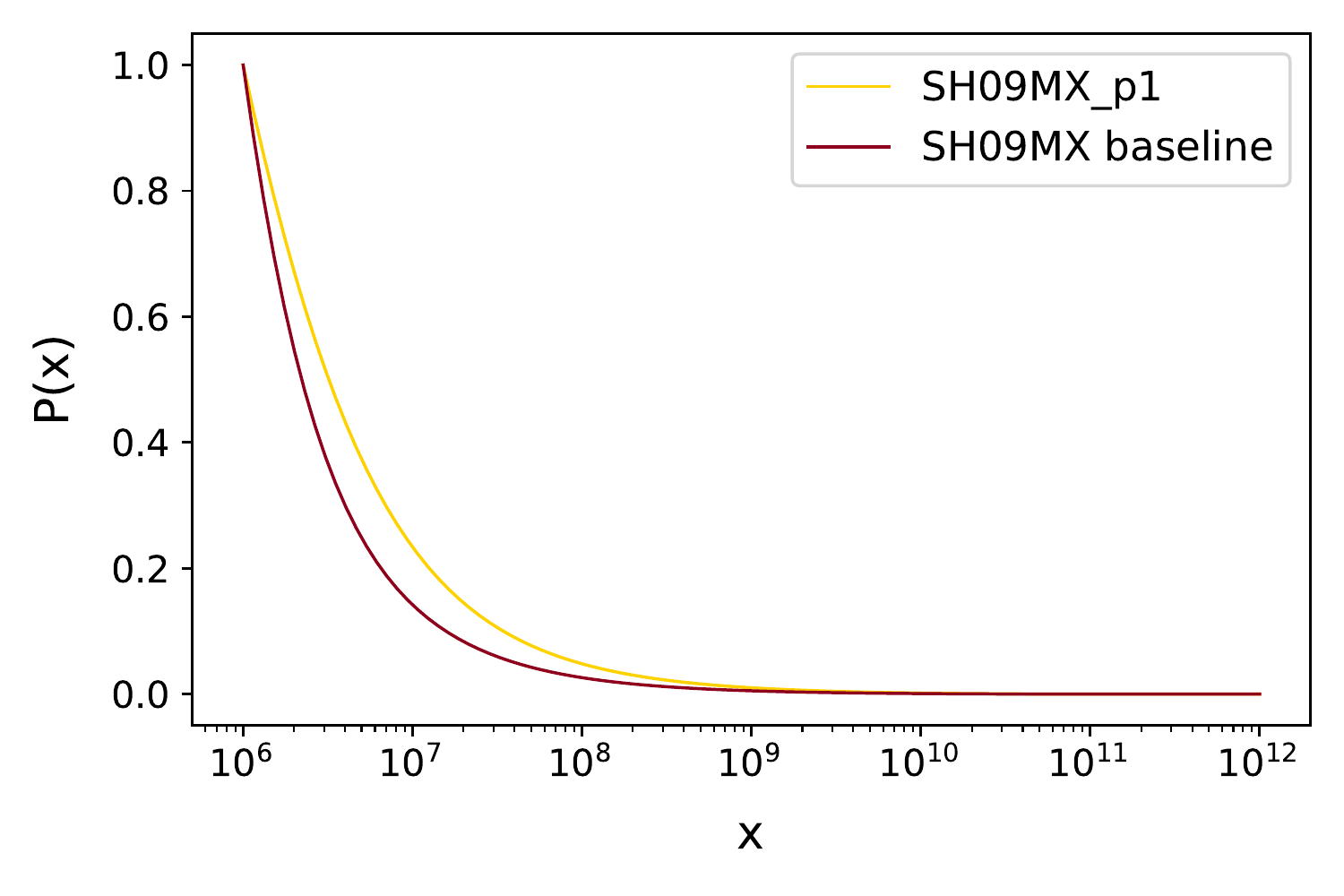}
    \includegraphics[width=\columnwidth,keepaspectratio]{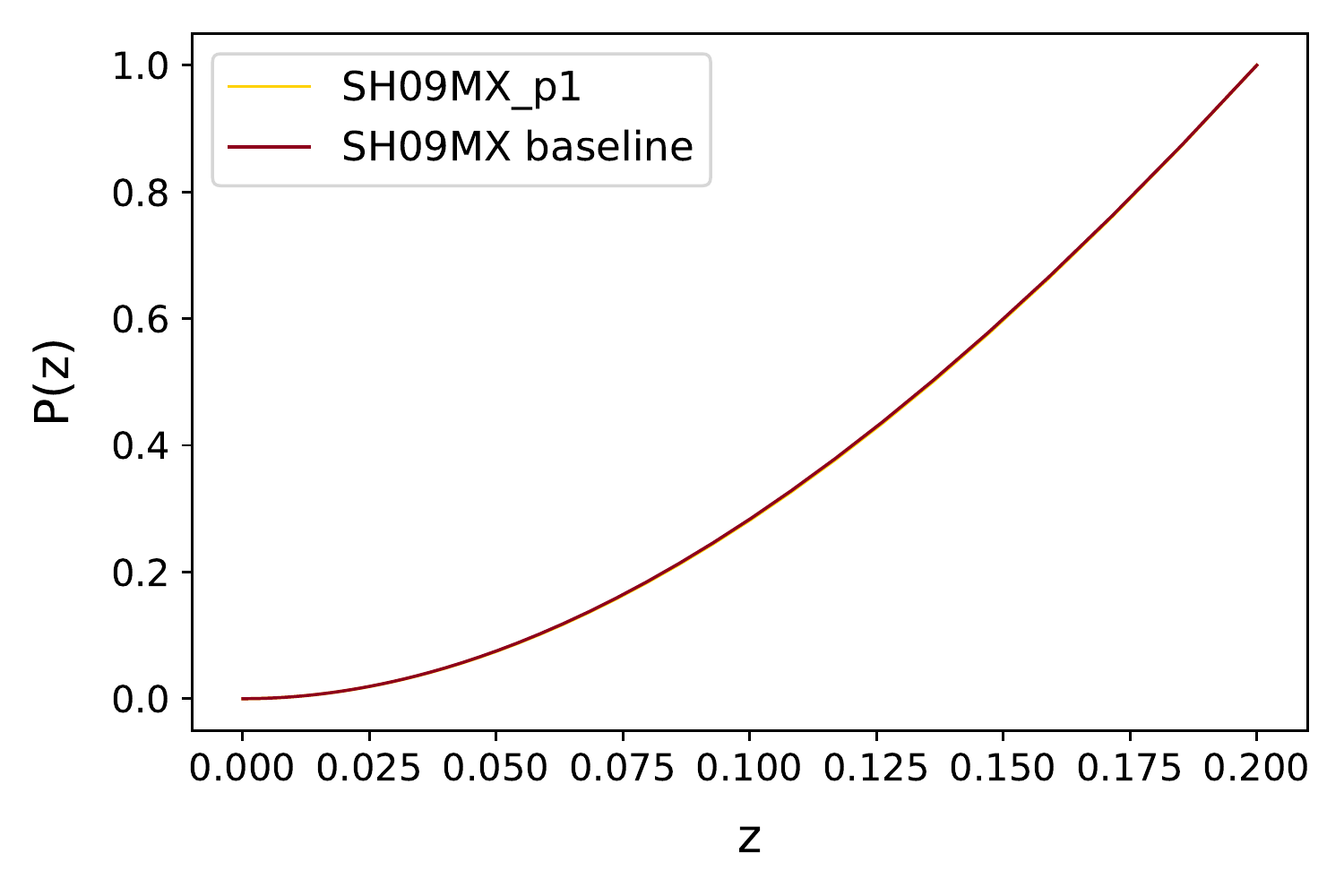}
    \caption{Marginalised 1D BHMF distributions for log-mass (\textbf{left}) and redshift (\textbf{right}) obtained from appropriately integrating the full BHMF in \autoref{eq.BHMF} for two different scalings of $P(x|z)$. }
    \label{fig:marginalised_distribution}
\end{figure*}

\begin{figure}
    \centering
    \includegraphics[width=\columnwidth,keepaspectratio]{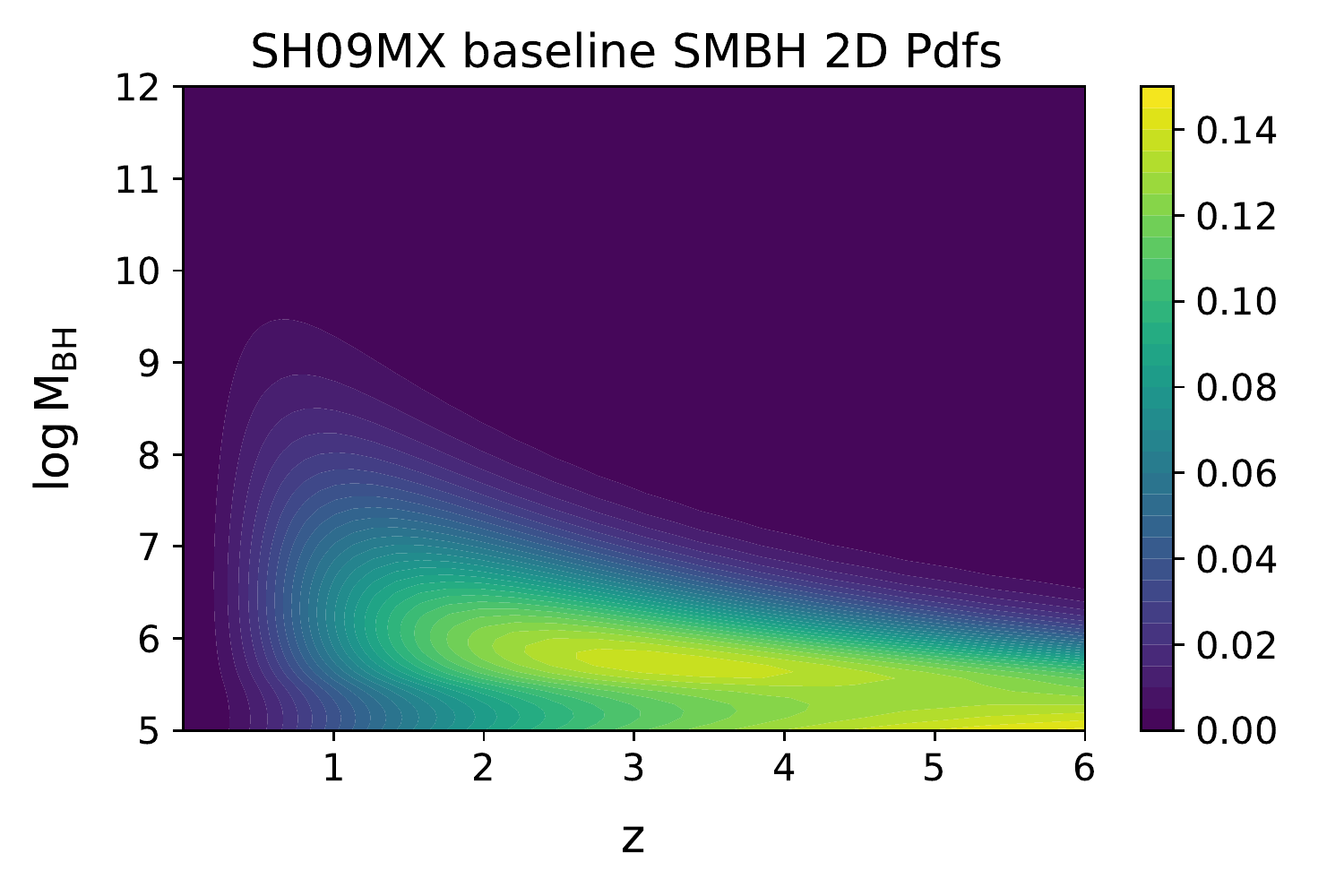}
    \includegraphics[width=\columnwidth,keepaspectratio]{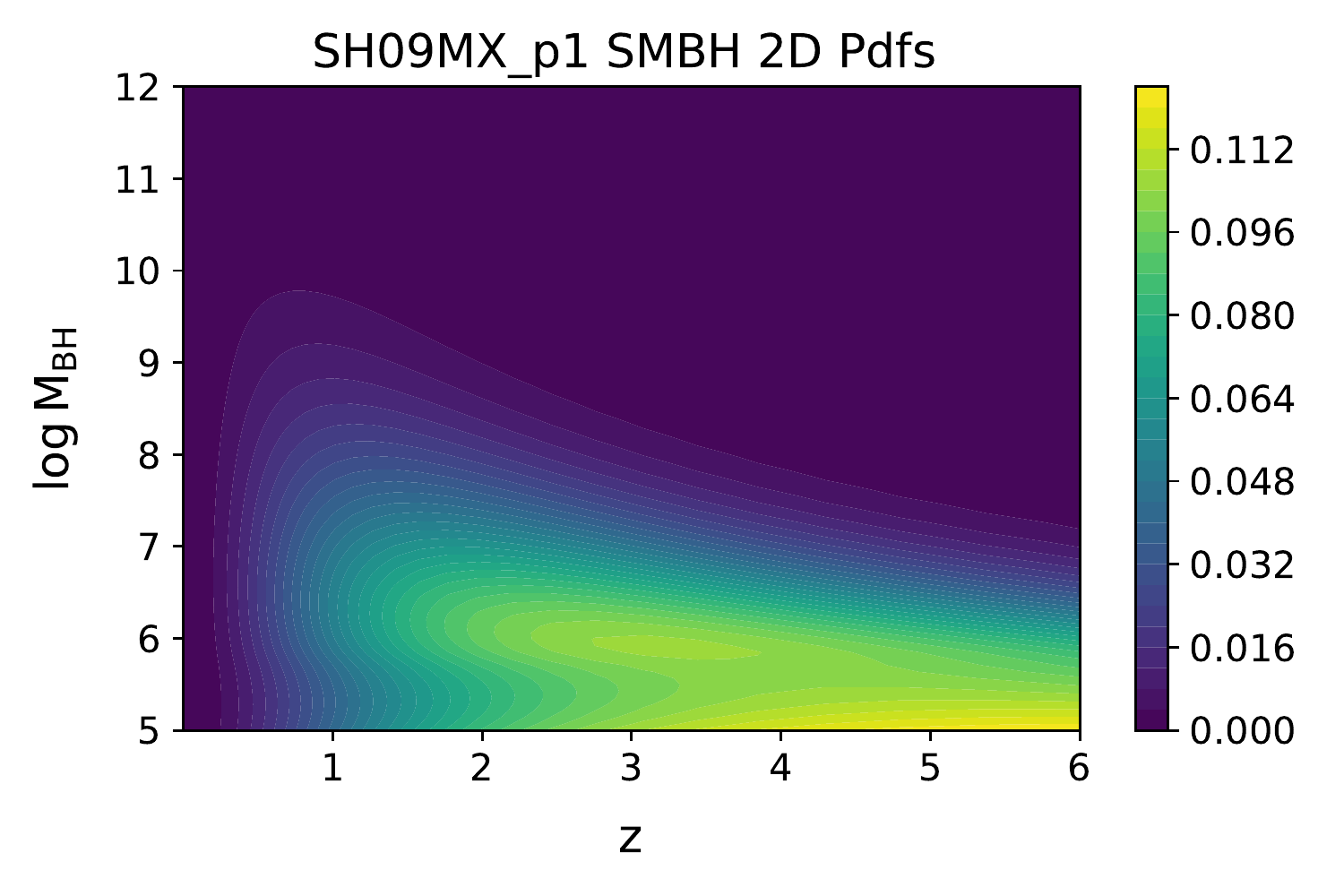}
    \caption{The BHMF represented in the plane $z$-$\log M_{\rm BH}$ for two different choices of the BHMF corresponding to $p = -1$ (\textbf{top}) and $p=-0.8$ (\textbf{bottom}), see also \autoref{eq.condP_scaling}. In both plots we report the BHMF divided into 30 isodensity contours.  }
    \label{fig:P_of_x_z}
\end{figure}

The corresponding marginalised distributions obtained with the two values of $p$ are shown in \autoref{fig:marginalised_distribution}.  
As we can appreciate from the plots, the change in the value of $p$ significantly impact the behaviour of $P(x)$ for SMBH masses $ \lesssim 10^9$ but it has a negligible impact for masses above this threshold. The reason is that the ultra massive BH in the tail of the distribution are exponentially less likely than those with smaller masses due to the behaviour of the BHMF we have chosen and this sums up to the imposed scaling of $P(x|z)$ leading to a fast depauperation of the high mass tail of $P(x)$ as the redshift increases (see also \autoref{fig:P_of_x_z}). However it must be taken into account that $P(x)$ is integrated over redshift and, therefore, the difference between the two curves accumulates as the redshift increases. 
Looking at the full 2D distribution (\autoref{fig:P_of_x_z}), it is clear that at low redshift the scaling of the conditional distribution has little impact, as its variance only decreases linearly with $z$. Thus, for the redshift of interest for the construction of the distance ladder, the choice of $P(x|z)$ is negligible even though it has a significant impact on the full 2D distribution. 

Furthermore, the impact on the redshift distribution is also negligible for the redshifts that are interesting for the distance ladder (i.e. $z \sim 0 $). Here the reason is that $P(z)$ depends on the size of the comoving volume and not only on the scaling of $P(x|z)$. For very small redshifts, $P(x|z)$ is almost a constant in $z$, therefore for $z \sim 0$ the scaling is determined solely by the scaling of the comoving volume, i.e. $dV/dz \sim \chi^2(z)/H(z)$. Therefore, there is no appreciable effect varying $p$ on the SMBH redshift distribution at low redshift.

In conclusion the distribution of SMBH at low redshift in our catalogue is determined by the chosen BHMF and the assumed cosmological model. We note however that choosing an accurate scaling for $P(x|z)$ and an accurate form of the BHMF is mandatory if one want to use SMBH shadows to reconstruct the distance-redshift relation at arbitrary redshift and not only to make the distance ladder.

\section{Spinning SMBHs and their shadows}\label{app.KerrBHs}

The analysis performed in this paper relies on the assumption that the observed SMBHs have a vanishing spin and can be described by a Schwarzschild metric, leading to the relation for the shadow size of \autoref{eq:angsize}.

However SMBHs do rotate and the effect of their spin on the surrounding environment alters the shape and the size of their shadows, quantities that now need to be computed in the Kerr metric, describing rotating BHs \cite{Li:2020drn,Takahashi:2004xh,Perlick:2017fio}. This is the case e.g. for the two SMBHs for which the shadow sizes have been measured by the EHT collaboration \cite{EventHorizonTelescope:2022xnr,EventHorizonTelescope:2022xqj,EventHorizonTelescope:2019pgp,EventHorizonTelescope:2019ggy}. 

It is therefore important to assess the impact of rotation on the constraints obtained in this work. 
Generally speaking, both the BH spin and the observer inclination angle will affect the size of the shadow, with departures from \autoref{eq:angsize} being maximal for an observer on the equatorial plane looking at a nearly-maximally spinning BH. In such a case, the shadow will be deformed of $\sim 7\%$ along the horizontal axis while preserving the same size of a Schwarzschild shadow on the vertical axis (these axis are defined in the plane perpendicular to the line of sight) \cite{Dymnikova_1986,Johannsen:2010ru,chan2013:abc}. 

Therefore, rotating BHs will exhibit an asymmetric shadow, with a degree of asymmetry determined by the value of the BH spin and observer inclination \cite{Perlick:2021aok}. Such an asymmetry, while complicating the description of the shadow size in terms of the BH parameters, allows to constrain the BH spin and observer inclination angle at the same time, independently of the mass of the BH \cite{Tamburini:2011tk,Tamburini:2019vrf}. 

In principle, one can use the asymmetry between the vertical and horizontal diameter of the shadow to improve constraints on the distance measure (see e.g. \cite{Perlick:2021aok} and reference therein). However, while such quantities are well defined theoretically, on the observational side it is very complicated to infer these two lengths, as the centroid of the shadow cannot be determined accurately from observations.
In order to overcome this issue, one can make use of quantities averaged over the full extension of the shadow,
which are well connected to what EHT-like interferometers would measure. One example is the average radius of the shadow, defined as \cite{Johannsen:2010ru,chan2013:abc}
\begin{equation}
    \langle  R\rangle \equiv \frac{1}{2\pi}\int^{2\pi}_{0} Rd\alpha\,.
\end{equation}

We note that analytical solutions for the shadow size in a Kerr space-time can be found only for a handful of cases \cite{Perlick:2021aok}, with none available for a generic spin and inclinations. 
Therefore, one needs to rely on phenomenological relations, obtained from numerical simulations of light rays travelling around a Kerr BH to determine the average size of the shadow. 
With this approach, the average radius (for an observer at distance much bigger than the horizon size) can be obtained in the form \cite{chan2013:abc}:

\begin{equation}
   \langle  R\rangle = R_0 + R_1\cos(2.14\, \theta_i - 22.9^{\circ})\,, \label{eq.fitting_avg_rad}
\end{equation}
with $R_0$ and $R_1$ being
\begin{align}
    R_0 &= (5.2 - 0.209a + 0.445a^2 - 0.567a^3 )\,m\,,  \notag\\
    R_1 &= 10^{-3}\left(0.24 - \frac{3.3}{(a-0.9017)^2 + 0.059}\right)\,m\,,\notag
\end{align}

where $a = J/(cM_{\rm BH})$, $J$ being the angular momentum of the BH, and $\theta_i$ being the angle between the observer line-of-sight and $J$. 

To understand the effect of spin and inclination in the determination of the Hubble constant, we introduce an additional step in the analysis proposed in the main text.
Instead of fitting directly to obtain $H_0$ from distances obtained with \autoref{eq:angsize}, we now need to first determine the distance comparing \autoref{eq.fitting_avg_rad} to the measured size of the shadow varying the mass, the spin, the inclination and the distance as free parameters; we can then convert the distance obtained this way into a constraint on $H_0$. This procedure must be done singularly for each SMBH in our catalogue, as it is done e.g. to determine the luminosity distance from gravitational wave events \cite{LIGOScientific:2019zcs}. For the purpose of showing the impact of spin and inclination on the constraints on $H_0$ we apply this exercise to the EHT-like case we proposed in the main text.

\begin{figure}
    \centering
    \includegraphics[width=0.99\columnwidth,keepaspectratio]{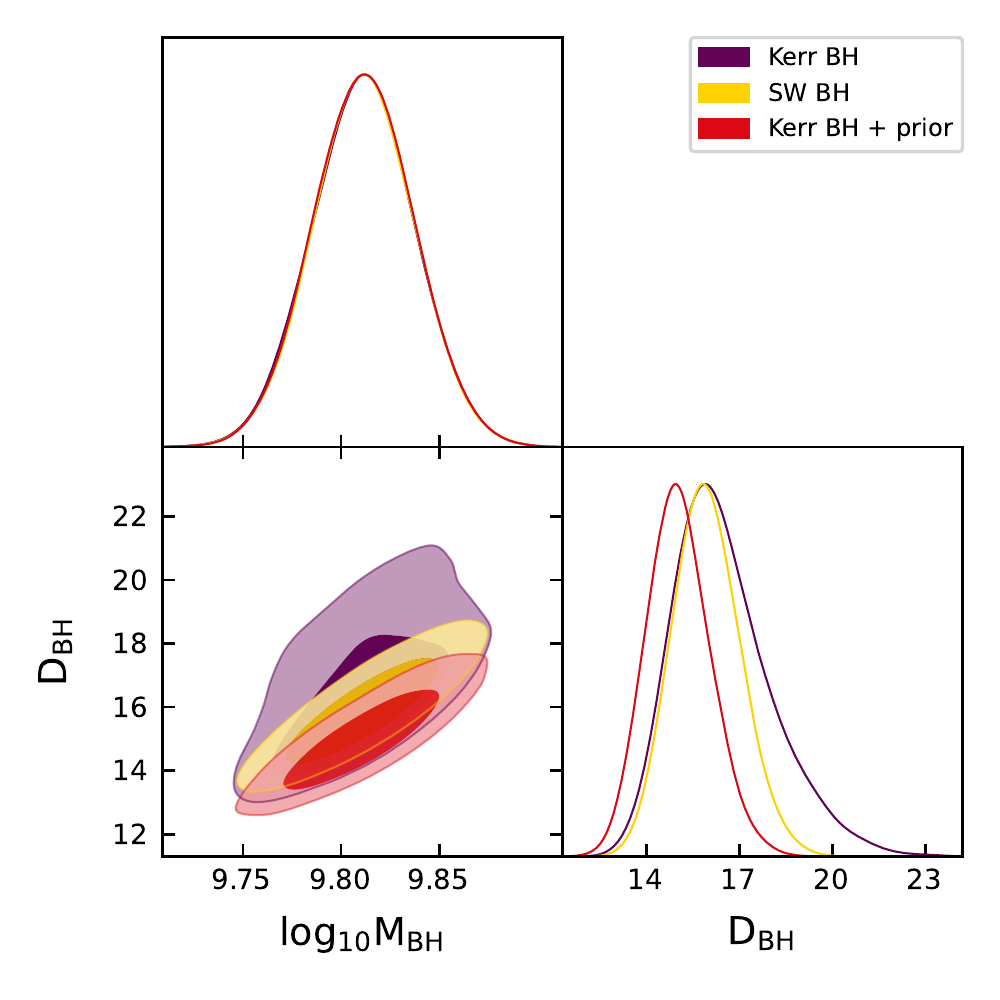}
    \caption{68$\%$\ and 95$\%$ C.L. on the distance and mass of the BH for three limit cases described in \ref{app.KerrBHs}}
    \label{fig:spin_params}
\end{figure}

The results are presented in \autoref{fig:spin_params}, where we show three limit cases for illustration:
\begin{enumerate}
    \item a general situation where only the mass of the BH is known while the spin and inclination are free to vary.
    We choose broad priors for these parameters namely $a \in [-1,1]$ and $\theta_i \in [0^{\circ},90^{\circ}]$.
    We label this case Kerr BH (purple contours);
    \item same as the previous case, but assuming the BH is not spinning ($a=0$), which corresponds to our baseline case where the shadow size is determined using \autoref{eq.BH_apparentsize}. We do however still use the inclination as a free parameter. We label this case SW BH (yellow contours);
    \item an optimal case where BH mass, BH spin and observer inclination are known (red contours). This case is exactly like the Kerr BH one, but with the inclusion of a prior on spin and inclination as determined in \cite{Tamburini:2019vrf} from the photon dynamic of the shadow of M87. We label this case as Kerr BH + prior.
\end{enumerate}

In all cases we assume the mass of the SMBH in M87 to be $M_{\rm BH}=(6.6 \pm 0.4)\cdot 10^9 $ M$_\odot$  \cite{Gebhardt:2011abc}.

To quantify the shift due to the inclusion of spin and inclination of the BH, we quantify the percentage error on the distance with the formula:
\begin{equation}
    \sigma_{\rm rel} = \frac{\sqrt{\sigma_{+}^2 + \sigma_{-}^2}}{\bar{\rm D}_{\rm BH}}
\end{equation}
where $\bar{\rm D}_{\rm BH}$ is the mean value of the posterior of the BH distance and $\sigma_{+,-}$ the right and left limit at 68$\%$ C.L. respectively . We find that $\sigma_{\rm rel} \sim 0.133$ for case (1) while $\sigma_{\rm rel} \sim 0.10$ for case (2) and (3).

\begin{figure}
    \centering
    \includegraphics[width=0.8\columnwidth]{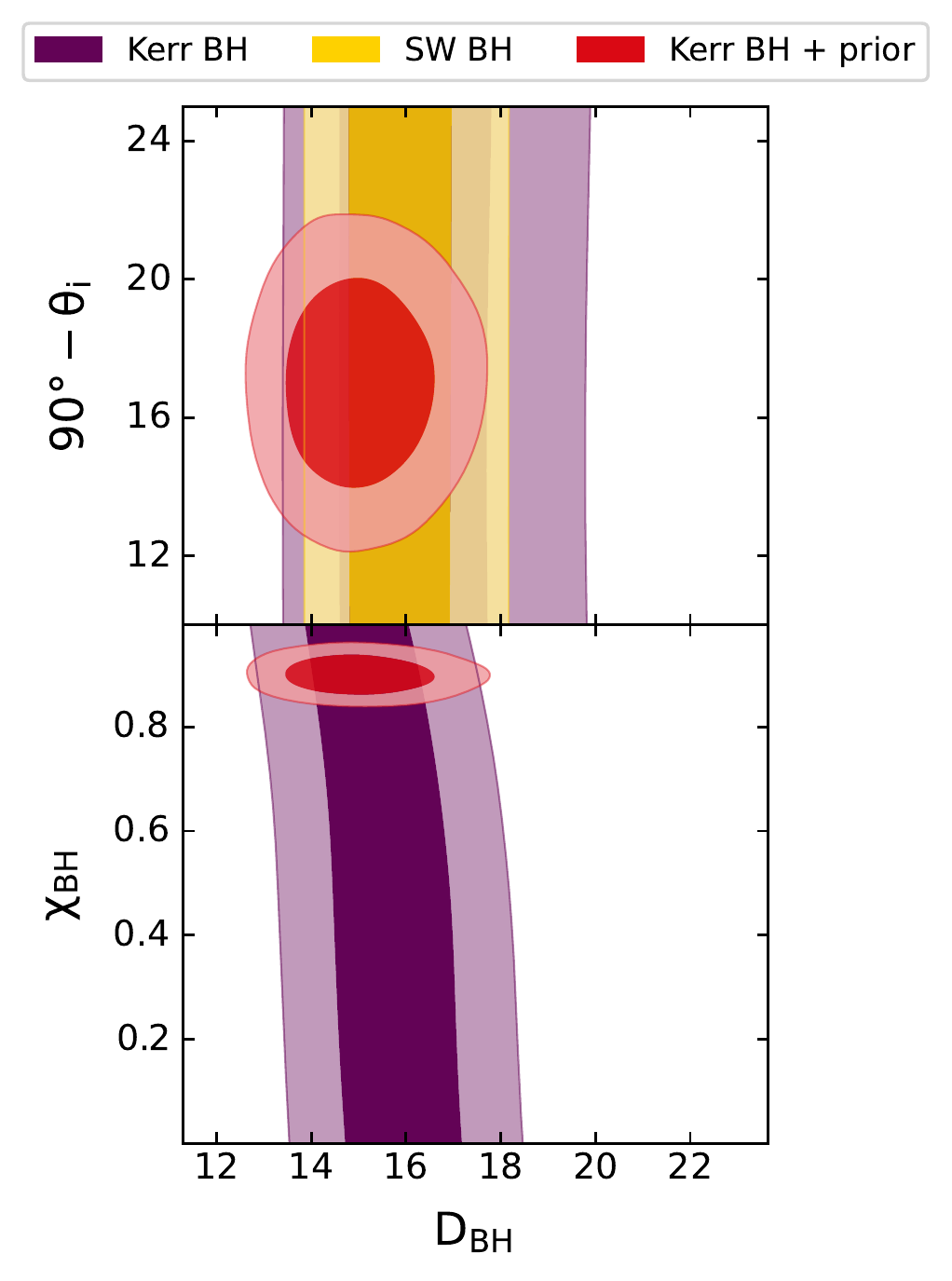}
    \caption{68$\%$\ and 95$\%$ C.L. on the distance, spin and inclination of the BH for three limit cases described in \ref{app.KerrBHs}. }
    \label{fig:D_corr}
\end{figure}

We further note that the spin and inclination parameters have a weak correlation with the determination of the BH distance when these parameters are free to vary. However, this correlation becomes more prominent for high spin, $|a| > 0.5$. This can be manifestly seen in the small shift of the distance posterior in the Kerr BH + prior case, see \autoref{fig:D_corr}. The uncertainty on the distance however stays unmodified compared with the SW BH case.

To determine the impact on the value of the Hubble constant we include this additional scatter in the EHT-like mock proposed in the main text. We found that the value of the Hubble constant is minimally affected leading to $H_0 =  69.8^{+6.6}_{-8.1} $ km s$^{-1}$ Mpc$^{-1}$, an increase of $\lesssim 1\%$ of the uncertainty.

We further note that due to the (small) correlation between spin, distance and mass of the BH, an increase in the uncertainty of the mass determination would reduce the impact of spin and inclination on the final errors, as the uncertainty on the mass dominates. As an example, a factor two worsening of the uncertainty on $M_{\rm BH}$ would lead to a $2\%$ error on the distance instead of the $3.5\%$ we find with our baseline $M_{\rm BH}$ uncertainty. On the contrary, should the mass measurement improve significantly, the inclusion of spin and inclination in the analysis would become crucial.


However given the current observational uncertainties, the impact of the BH spin and observer inclination has almost no influence on the measurement of distances from SMBH shadows. These parameters, in fact, increase the uncertainty on the distance measure only by $\sim 3.5\%$ ($\lesssim 1\%$ on $H_0$) if their values are completely unknown. Nevertheless, given that those parameters can be determined by the dynamic of the accretion disk around the BH \cite{Tamburini:2019vrf,Wielgus:2021peu}, with some prior knowledge of the spin and inclination parameters it is possible to achieve the same accuracy obtained with \autoref{eq.BH_apparentsize} on $\rm D_{\rm BH}$.

Overall, our analysis hints to the fact rotating BHs do not pose significant limitation to the use of SMBH shadows as cosmological probe, at least with the current level of observational uncertainty. 
The dominant contribution to this measurement is still given by the determination of the mass (as well as the shadow size) even in the case of Kerr BHs. 




%
%

\section{On the mass determination of cosmological SMBHs}
\label{app:uncertainties}


Throughout the analysis done in this paper, even when varying the observational specifications such as the value of $\theta_{\rm cut}$, we have kept fixed the uncertainties on the measurements of $\theta_{\rm BH}$ and $M_{\rm BH}$ to the level of $7\%$. This was motivated by existing observations (see \autoref{sec:datasets}), but such uncertainties are indeed optimistic when dealing with observations of shadows at distances higher than those currently observed.

In particular, there are no reliable and robust methods that can be applied to every system to infer the mass of SMBH, a significant (and arguably the most prominent) limitation to the use of these objects as distance tracers. The main issue is that the available methods are precise, but not necessarily accurate, an indication that astrophysical systematics are still poorly understood. As an example, $\rm M87^\star$ has two mass determinations to date: stellar kinematics gives $M_{\rm BH}=(6.6 \pm 0.4)\cdot 10^9 $ M$_\odot$ \cite{Gebhardt:2011abc} while gas dynamics gives $M_{\rm BH}=(3.5^{+0.9}_{-0.7})\cdot 10^9 $ M$_\odot$\cite{Walsh:2013uua}. The two determinations have an accuracy of $\sim 7\%$ and $\sim 22\%$ respectively, but they exhibit a $\sim 3.5\sigma$ tension showing that systematic uncertainties are still far from being under control. The EHT collaboration has also determined the mass from the shadow size of M87 to be $M_{\rm BH}= 6.5 \pm 0.2|_{\rm stat} \pm 0.7|_{\rm sys} \cdot 10^9$ M$_\odot$ \cite{EventHorizonTelescope:2019ggy} showing good agreement with the  measurement of \cite{Gebhardt:2011abc} and a $3 \sigma$ discrepancy with that of \cite{Walsh:2013uua}. 

In order to avoid this issue, one could rely on more conservative estimates of the observed quantities ($\theta_{\rm BH}$ and $M_{\rm BH}$), in order to avoid inaccuracies due to systematic effects. This however increases the observational uncertainties on the distance to the BH, and will impact the final constraints on $H_0$. In order to estimate how the final constraints degrade with increased uncertainties, we performed again our baseline analysis, but this time we vary, separately, the relative uncertainties $\sigma_\theta/\theta_{\rm BH}$ and $\sigma_M/M_{\rm BH}$ in the range $[7\%,80\%]$.

\begin{figure}
    \centering
    \includegraphics[width=.9\columnwidth,keepaspectratio]{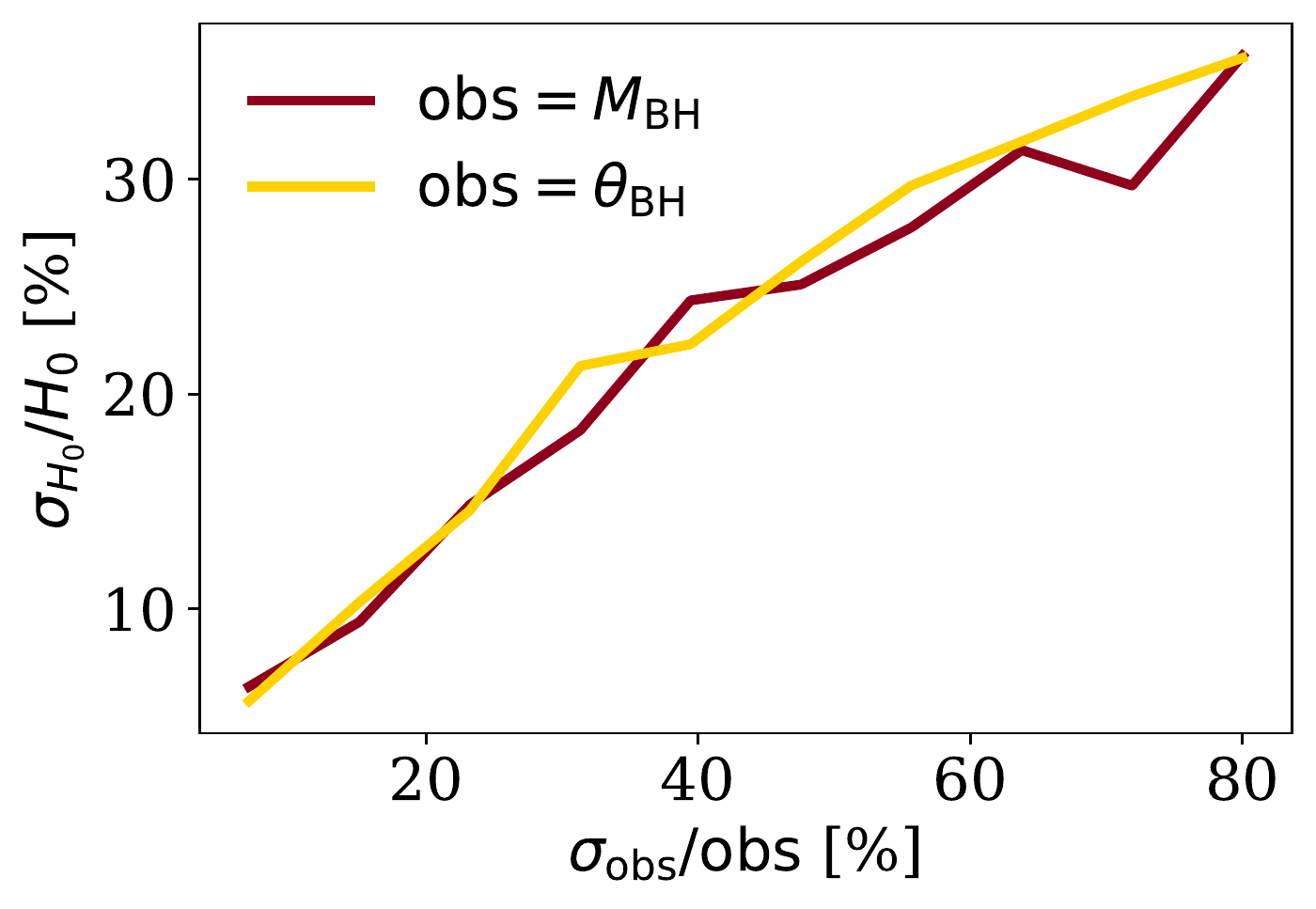}
    \caption{Trend of the relative error on $H_0$ (in percentage) with changes of the percentage error on the observed size of the shadow $\theta_{\rm BH}$ (yellow line) and on the BH mass $M_{\rm BH}$ (red line).}
    \label{fig:uncertainties}
\end{figure}

Our results are shown in \autoref{fig:uncertainties}, yellow and red lines showing how changes in $\sigma_\theta/\theta_{\rm BH}$ and $\sigma_M/M_{\rm BH}$, respectively, translate into changes in the relative error on $H_0$. We find that indeed the precision with which the observables are measured have a significant impact, with the final constraint on $H_0$ ranging from $\approx5\%$ to $\approx40\%$ moving from the most optimistic to the most pessimistic case.

\bibliographystyle{elsarticle-num-names}
\bibliography{bibliography}

\end{document}